\documentclass[12pt]{article}

\usepackage{epsfig}
\usepackage{color}
\usepackage{array}
\usepackage{amsmath}
\usepackage{amssymb}
\usepackage{bm}

\newcommand{\bea}{\begin{eqnarray}}
\newcommand{\eea}{\end{eqnarray}}

 \makeatletter
\def\alt{\mathrel{\mathpalette\gl@align<}}
\def\agt{\mathrel{\mathpalette\gl@align>}}
\def\gl@align#1#2{\lower.6ex\vbox{\baselineskip\z@skip\lineskip\z@
\ialign{$\m@th#1\hfil##\hfil$\crcr#2\crcr\sim\crcr}}} \makeatother

 \makeatletter
\def\alt{\mathrel{\mathpalette\gl@align<}}
\def\agt{\mathrel{\mathpalette\gl@align>}}
\def\gl@align#1#2{\lower.6ex\vbox{\baselineskip\z@skip\lineskip\z@
\ialign{$\m@th#1\hfil##\hfil$\crcr#2\crcr\sim\crcr}}} \makeatother

\catcode`@=11
\long\def\@caption#1[#2]#3{\par\addcontentsline{\csname
  ext@#1\endcsname}{#1}{\protect\numberline{\csname
  the#1\endcsname}{\ignorespaces #2}}\begingroup
    \small
    \@parboxrestore
    \@makecaption{\csname fnum@#1\endcsname}{\ignorespaces #3}\par
  \endgroup}
\catcode`@=12

\begin{document}
\begin{flushright}
OSU-HEP-10-04\\
June 2010
\end{flushright}
\vspace*{2.5cm}

\begin{center}
\baselineskip 20pt {\Large\bf Semidirect Product Groups,
Vacuum Alignment\\[0.05in]
and Tribimaximal Neutrino Mixing}
\vspace{1.0cm}

{\large \it K.S. Babu\,\footnote{Email: babu@okstate.edu}} and {\large\it S. Gabriel\,\footnote{svengab@msn.com}}

\vspace{0.2cm}

{\it Department of Physics, Oklahoma State University, Stillwater, OK 74078, USA}

\begin{abstract}
The neutrino oscillation data are in very good agreement with the tribimaximal
mixing pattern: $\sin^2\theta_{23}=1/2$, $\sin^2\theta_{12}=1/3$, and
$\sin^2\theta_{13}=0$.  Attempts to generate this pattern based on
finite family symmetry groups typically assume that the family symmetry
is broken to different subgroups in the charged lepton and the neutrino
mass matrices. This leads to a technical problem, where the cross-couplings between
the Higgs fields responsible for the two symmetry breaking chains force their vacuum
expectation values to align, upsetting the desired breaking pattern. Here,
we present a class of models based on the semidirect product group $(S_3)^4 \rtimes A_4$,
where the lepton families belong to representations which are not
faithful. In effect, the Higgs sector knows about the full symmetry while the
lepton sector knows only about the $A_4$ factor group. This can solve the alignment
problem without altering the desired properties of the family symmetry.
Inclusion of quarks into the framework is straightforward, and leads to small
and arbitrary CKM mixing angles. Supersymmetry is not essential for our proposal, but
the model presented is easily supersymmetrized, in which case the same family symmetry
solves the SUSY flavor problem.
\end{abstract}

\end{center}

\newpage

\addtocounter{page}{-1}

\section{Introduction}

The neutrino mixing angles are now known to some accuracy \cite{schwetz,fogli}. A global analysis of
data shows (in the standard parametrization of the lepton mixing matrix)  \cite{schwetz}:
\begin{equation}
\label{data}
\sin^2\theta_{23}=0.50^{+0.07}_{-0.06}\,,~~
\sin^2\theta_{12}=0.304^{+0.022}_{-0.016}\,,~~ \sin^2\theta_{13}<0.056\,,\end{equation}
where the error bars correspond to $1 \sigma$ variations and the upper limit is at the $3 \sigma$ level.
These values are in good agreement with the tribimaximal mixing
pattern given by the leptonic mixing matrix [3, 4]
\begin{eqnarray}U_{MNS}=
\label{tribi}
\left(
                          \begin{array}{ccc}
                            \sqrt{\frac{2}{3}} & \frac{1}{\sqrt{3}} & 0 \\
                            -\frac{1}{\sqrt{6}} & \frac{1}{\sqrt{3}} & -\frac{1}{\sqrt{2}} \\
                            -\frac{1}{\sqrt{6}} & \frac{1}{\sqrt{3}} & \frac{1}{\sqrt{2}} \\
                          \end{array}
                        \right)P
\end{eqnarray} where $P$ is a diagonal phase matrix. This matrix corresponds to geometric
values of the mixing parameters
\begin{equation}
\sin^2\theta_{23}=1/2,~~\sin^2\theta_{12}=1/3,~~\sin^2\theta_{13}=0\,\end{equation}
which are all in good agreement presently with the experimental values quoted in (\ref{data}).

It has been known for some time that the tribimaximal mixing pattern of (\ref{tribi}) can be obtained
using finite family symmetry groups such as $A_4$ \cite{ma-rajasekaran,ma2002,babu-ma-valle,altarelli,babu-he,he-keum-volkas,ma2009,hmv,fhm,af2010},
and others \cite{other}. In
the class of models based on $A_4$, the standard model (SM) lepton doublets are assigned to triplet representation of $A_4$, and the $A_4$ symmetry is broken to a $Z_2$ subgroup in the neutrino
sector by a triplet Higgs $\chi$, with the vacuum expectation value (VEV) structure $\left\langle \chi \right\rangle  \propto (0,1,0)$ or some permutation thereof, and to a $Z_3$ subgroup in the charged lepton
sector by a triplet Higgs $\phi$, with the VEV structure $\left\langle \phi \right\rangle \propto (1,1,1)$.
However, there is a technical problem with this, in that couplings between the Higgs fields $\chi$ and $\phi$ will force their VEVs to align, upsetting the desired breaking pattern.  Several approaches have been adopted in the
literature to overcome this vacuum alignment problem. One can introduce more complicated symmetries, and assume that some symmetries are softly broken.  Alternatively, one can assume that the symmetry is only approximate, and that non-renormalizable operators are present in the Lagrangian.  Supersymmetry helps, since the alignment
problem exists in this context only at the cubic level in the superpotential. Care must be taken to ensure
that there are no flat directions -- a frequent occurrence in SUSY models.  
The existence of extra dimensions is sometimes assumed,
where the two Higgs fields can be separated, thus circumventing the alignment problem.
While all of these approaches are interesting, in our opinion, a compelling scenario has not emerged, arising from renormalizable models with exact symmetries realized in the Higgs phase.

In this paper we propose a new approach to address the vacuum alignment problem
using semidirect product groups of a type usually not considered,
which provides a technically complete picture for tribimaximal
lepton mixing.  Supersymmetry is not essential, although the idea
presented is compatible with SUSY.  The symmetries of the model are
assumed to be spontaneously (and not softly) broken.
Non-renormalizable operators are not necessary, although their
existence does not upset the desired VEV structure.

We consider models where the SM lepton families belong to
representations of a finite symmetry which are not faithful (that
is, not every member of the group is represented by a distinct
transformation). In effect, the Higgs sector knows about the full
symmetry while the lepton sector does not. Our framework is a
renormalizable extension of the standard model with an additional
finite family symmetry $G\times Z_2$, with $G=S_3^4\rtimes A_4$. The
rationale for the choice of this group is elaborated in Sec. II.
Here we note briefly that $S_3$ is the smallest non-Abelian group,
and that four such factors would allow one to evenly permute them,
which is the symmetry operation of $A_4$.  This enables us to form a
non-trivial semidirect product group, which has all the desired
properties for solving the vacuum alignment problem. This group
contains $S_3^4$ as an invariant subgroup, so that it will have
representations corresponding to the factor group $G/S_3^4\sim A_4$.
The standard model leptons can then be assigned to representations
of $A_4$, as has been done in the models of Ref.
\cite{ma-rajasekaran,ma2002,babu-ma-valle,altarelli,babu-he,he-keum-volkas,ma2009,hmv,fhm,af2010}.
Neutrino masses are generated by a Higgs field $\phi$, belonging to
a $Z_2$ even $16$-dimensional representation of $G$, while
charged-lepton masses are generated by a Higgs field $\chi$,
belonging to a $Z_2$ odd triplet of $A_4$. The $Z_2$ symmetry
prevents cubic and non-trivial quartic interactions between $\phi$
and $\chi$. In this way, the alignment problem is solved without
altering the desired properties of the family symmetry. Inclusion of
the quark sector into the framework is straightforward, and leads to
arbitrary and small CKM mixing angles. The model can readily be
supersymmetrized, as discussed in Sec. VI. In this case the same
family symmetry can solve the SUSY flavor problem.

In Sec. II we present our model based on $(S_3^4 \rtimes A_4) \times
Z_2$ family symmetry.  In Sec. III the Higgs potential of the model
is analyzed, where it is established that the model has no vacuum
alignment problem. Here we also assert, by computing the scalar
boson masses, that the symmetry breaking is complete, without
leading to any pseudo-Goldston bosons.  Sec. IV discusses the lepton
mass matrices, and the resulting tribimaximal lepton mixing matrix.  In Sec. V
we discuss briefly the inclusion of quarks in the framework.  Sec. VI discusses
supersymmetrization of the model. In Sec. VII we conclude.  In four
Appendices we provide various technical details:  the generators of
the group in matrix form, invariants of the group we need,
and the calculations of the light charged lepton and neutrino
mass matrices.

\section{Model of $(S_3^4 \rtimes A_4) \times Z_2$ family symmetry}

The gauge symmetry of our model is $SU(3)_c \times SU(2)_L \times U(1)_Y$, and we
work in its minimal non-supersymmetric version.
We extend the gauge symmetry by the finite family symmetry $G\times Z_2$, where $G$ has the
semi-direct product\footnote{The semi-direct product, $N\rtimes H$,
contains $N$ and $H$ as subgroups and obeys $hnh^{-1}\in N$ for all
$n\in N$ and $h\in H$ \cite{kim}. Thus, $N$ is an invariant
subgroup. The number of elements in the group, denoted by $|N\rtimes
H|$, is $|N||H|$. The semi-direct product exists when $H$ has a
factor group which is a subgroup of the automorphism group of $N$.}
structure $(S_3\times S_3\times S_3\times S_3)\rtimes A_4$. The
finite symmetry is broken at a scale $M_*$ which is assumed to be large compared
to the EW scale. $M_*$ may be identified as the grand unification scale,
or it may be an intermediate scale.

\subsection{Symmetry group}

The group $A_4$ can be described using two
generators obeying the relations,
\begin{eqnarray}X^2=Y^3=E,~~XYX=Y^2XY^2,\label{group1}\end{eqnarray}
where $E$ is the identity. The irreducible representations of $A_4$ are one
real singlet, two complex singlets, and one real triplet. Table 1
gives $X$ and $Y$ in each of these representations. The $S_3$
generators, $A_i$ and $B_i$ obey
\begin{eqnarray}A_i^3=B_i^2=E,~~~~B_iA_iB_i^{-1}=A_i^{-1}.\end{eqnarray}
The irreducible representations of $S_3$ are two real singlets
and one real doublet. The generators in each of these
representations can also be found in Table 1. The remaining
relations defining the full symmetry are
\begin{eqnarray}XA_1X^{-1}=A_2,~~~~XA_2X^{-1}=A_1,~~~~XA_3X^{-1}=A_4,~~~~XA_4X^{-1}=A_3,\nonumber\end{eqnarray}
\vspace{-1cm}
\begin{eqnarray}XB_1X^{-1}=B_2,~~~~XB_2X^{-1}=B_1,~~~~XB_3X^{-1}=B_4,~~~~XB_4X^{-1}=B_3,\end{eqnarray}
\begin{eqnarray}YA_1Y^{-1}=A_1,~~~~YA_2Y^{-1}=A_3,~~~~YA_3Y^{-1}=A_4,~~~~YA_4Y^{-1}=A_2,\nonumber\end{eqnarray}
\vspace{-1cm}
\begin{eqnarray}YB_1Y^{-1}=B_1,~~~~YB_2Y^{-1}=B_3,~~~~YB_3Y^{-1}=B_4,~~~~YB_4Y^{-1}=B_2.
\label{group4}
\end{eqnarray}
\begin{table}
\begin{center}
\begin{tabular}{|c|c|c|}
  \hline
  ~ & X & Y \\\hline
  $1$ & 1 & 1 \\
  $1'$ & 1 & $\omega$ \\
  $1''$ & 1 & $\omega^2$ \\
  $3$ & $\left(
        \begin{array}{ccc}
          -1 & 0 & 0 \\
          0 & 1 & 0 \\
          0 & 0 & -1 \\
        \end{array}
      \right)$
   & $\left(
       \begin{array}{ccc}
         0 & 0 & 1 \\
         1 & 0 & 0 \\
         0 & 1 & 0 \\
       \end{array}
     \right)$
    \\\hline
\end{tabular}\hspace{1cm}\begin{tabular}{|c|c|c|}
                 \hline
                 $~$ & $A_i$ & $B_i$ \\\hline
                 1 & 1 & 1 \\
                 $1'$ & 1 & -1 \\
                 2 & $\left(
                       \begin{array}{cc}
                         \omega & 0 \\
                         0 & \omega^2 \\
                       \end{array}
                     \right)$
 & $\left(
                       \begin{array}{cc}
                         0 & 1 \\
                         1 & 0 \\
                       \end{array}
                     \right)$ \\
                 \hline
               \end{tabular}\\
\vspace{0.5cm}
\end{center}\caption{The left panel shows the matrices representing the generators in each irrep. of
$A_4$, in a certain basis. The right panel shows the same for
$S_3$. Here $\omega=e^{2i\pi/3}$.}
\end{table}It's easy to see that if $A_1$, $A_2$, $A_3$, $A_4$, $B_1$, $B_2$, $B_3$, and $B_4$ are
all represented by the identity matrix, then (5), (6), and (7) are
trivially satisfied, leaving only (4) to be checked. But finding
representations obeying (4) corresponds to finding the
representations of $A_4$, which are given in Table 1. These
representations are used for SM leptons.  The character table of
$S_3^4\rtimes A_4$ is given in Table 2 for completeness. The first
four representations listed in Table 2 correspond to the
representations of $A_4$.

\begin{table}[tbh]
\begin{center}
{\tiny\begin{tabular}{|c|c|c|c|c|c|c|c|c|c|c|c|c|c|c|c|c|c|c|c|c|c}
  \hline
  $~$ & 1 & 8 & 24 & 32 & 16 & 12 & 54 & 108 & 81 & 72 & 144 & 216 & 96 & 216 & 216 & 108 & 432 & 432 &
  648 & 1296 & $\dots$
  \\\hline
  1 & 1 & 1 & 1 & 1 & 1 & 1 & 1 & 1 & 1 & 1 & 1 & 1 & 1 & 1 & 1 & 1 & 1 & 1 & 1 & 1 & $\dots$ \\
  $1'$ & 1 & 1 & 1 & 1 & 1 & 1 & 1 & 1 & 1 & 1 & 1 & 1 & 1 & 1 & 1 & 1 & 1 & 1 & 1 & 1 & $\dots$ \\
  $1''$ & 1 & 1 & 1 & 1 & 1 & 1 & 1 & 1 & 1 & 1 & 1 & 1 & 1 & 1 & 1 & 1 & 1 & 1 & 1 & 1 & $\dots$ \\
  3 & 3 & 3 & 3 & 3 & 3 & 3 & 3 & 3 & 3 & 3 & 3 & 3 & 3 & 3 & 3 & -1 & -1 & -1 & -1 & -1 & $\dots$ \\
  $\widetilde{1}$ & 1 & 1 & 1 & 1 & 1 & -1 & 1 & -1 & 1 & -1 & -1 & 1 & -1 & 1 & -1 & 1 & 1 & 1 & -1 & -1 & $\dots$ \\
  $\widetilde{1}'$ & 1 & 1 & 1 & 1 & 1 & -1 & 1 & -1 & 1 & -1 & -1 & 1 & -1 & 1 & -1 & 1 & 1 & 1 & -1 & -1 & $\dots$ \\
  $\widetilde{1}''$ & 1 & 1 & 1 & 1 & 1 & -1 & 1 & -1 & 1 & -1 & -1 & 1 & -1 & 1 & -1 & 1 & 1 & 1 & -1 & -1 & $\dots$ \\
  $\widetilde{3}$ & 3 & 3 & 3 & 3 & 3 & -3 & 3 & -3 & 3 & -3 & -3 & 3 & -3 & 3 & -3 & -1 & -1 & -1 & 1 & 1 & $\dots$ \\
  4 & 4 & 4 & 4 & 4 & 4 & 2 & 0 & -2 & -4 & 2 & 2 & 0 & 2 & 0 & -2 & 0 & 0 & 0 & 0 & 0 & $\dots$ \\
  $4'$ & 4 & 4 & 4 & 4 & 4 & 2 & 0 & -2 & -4 & 2 & 2 & 0 & 2 & 0 & -2 & 0 & 0 & 0 & 0 & 0 & $\dots$ \\
  $4''$ & 4 & 4 & 4 & 4 & 4 & 2 & 0 & -2 & -4 & 2 & 2 & 0 & 2 & 0 & -2 & 0 & 0 & 0 & 0 & 0 & $\dots$ \\
  $\widetilde{4}$ & 4 & 4 & 4 & 4 & 4 & -2 & 0 & 2 & -4 & -2 & -2 & 0 & -2 & 0 & 2 & 0 & 0 & 0 & 0 & 0 & $\dots$ \\
  $\widetilde{4}'$ & 4 & 4 & 4 & 4 & 4 & -2 & 0 & 2 & -4 & -2 & -2 & 0 & -2 & 0 & 2 & 0 & 0 & 0 & 0 & 0 & $\dots$ \\
  $\widetilde{4}''$ & 4 & 4 & 4 & 4 & 4 & -2 & 0 & 2 & -4 & -2 & -2 & 0 & -2 & 0 & 2 & 0 & 0 & 0 & 0 & 0 & $\dots$ \\
  6 & 6 & 6 & 6 & 6 & 6 & 0 & -2 & 0 & 6 & 0 & 0 & -2 & 0 & -2 & 0 & 2 & 2 & 2 & 0 & 0 & $\dots$ \\
  $6'$ & 6 & 6 & 6 & 6 & 6 & 0 & -2 & 0 & 6 & 0 & 0 & -2 & 0 & -2 & 0 & -2 & -2 & -2 & 0 & 0 & $\dots$ \\
  8 & 8 & 5 & 2 & -1 & -4 & 6 & 4 & 2 & 0 & 3 & 0 & 1 & -3 & -2 & -1 & 0 & 0 & 0 & 0 & 0 & $\dots$ \\
  $8'$ & 8 & 5 & 2 & -1 & -4 & 6 & 4 & 2 & 0 & 3 & 0 & 1 & -3 & -2 & -1 & 0 & 0 & 0 & 0 & 0 & $\dots$ \\
  $8''$ & 8 & 5 & 2 & -1 & -4 & 6 & 4 & 2 & 0 & 3 & 0 & 1 & -3 & -2 & -1 & 0 & 0 & 0 & 0 & 0 & $\dots$ \\
  $\widetilde{8}$ & 8 & 5 & 2 & -1 & -4 & -6 & 4 & -2 & 0 & -3 & 0 & 1 & 3 & -2 & 1 & 0 & 0 & 0 & 0 & 0 & $\dots$ \\
  $\widetilde{8}'$ & 8 & 5 & 2 & -1 & -4 & -6 & 4 & -2 & 0 & -3 & 0 & 1 & 3 & -2 & 1 & 0 & 0 & 0 & 0 & 0 & $\dots$ \\
  $\widetilde{8}''$ & 8 & 5 & 2 & -1 & -4 & -6 & 4 & -2 & 0 & -3 & 0 & 1 & 3 & -2 & 1 & 0 & 0 & 0 & 0 & 0 & $\dots$ \\
  16 & 16 & -8 & 4 & -2 & 1 & 0 & 0 & 0 & 0 & 0 & 0 & 0 & 0 & 0 & 0 & 4 & -2 & 1 & 0 & 0 & $\dots$ \\
  $16'$ & 16 & -8 & 4 & -2 & 1 & 0 & 0 & 0 & 0 & 0 & 0 & 0 & 0 & 0 & 0 & 4 & -2 & 1 & 0 & 0 & $\dots$ \\
  $16''$ & 16 & -8 & 4 & -2 & 1 & 0 & 0 & 0 & 0 & 0 & 0 & 0 & 0 & 0 & 0 & 4 & -2 & 1 & 0 & 0 & $\dots$ \\
  24 & 24 & 6 & -3 & -3 & 6 & 12 & 4 & 0 & 0 & 0 & -3 & -2 & 3 & 1 & 0 & 4 & 1 & -2 & 2 & -1 & $\dots$ \\
  $24'$ & 24 & 6 & -3 & -3 & 6 & 12 & 4 & 0 & 0 & 0 & -3 & -2 & 3 & 1 & 0 & -4 & -1 & 2 & -2 & 1 & $\dots$ \\
  $\widetilde{24}$ & 24 & 6 & -3 & -3 & 6 & -12 & 4 & 0 & 0 & 0 & 3 & -2 & -3 & 1 & 0 & 4 & 1 & -2 & -2 & 1 & $\dots$ \\
  $\widetilde{24}'$ & 24 & 6 & -3 & -3 & 6 & -12 & 4 & 0 & 0 & 0 & 3 & -2 & -3 & 1 & 0 & -4 & -1 & 2 & 2 & -1 & $\dots$ \\
  $24''$ & 24 & 15 & 6 & -3 & -12 & 6 & -4 & -6 & 0 & 3 & 0 & -1 & -3 & 2 & 3 & 0 & 0 & 0 & 0 & 0 & $\dots$ \\
  $\widetilde{24}''$ & 24 & 15 & 6 & -3 & -12 & -6 & -4 & 6 & 0 & -3 & 0 & -1 & 3 & 2 & -3 & 0 & 0 & 0 & 0 & 0 & $\dots$ \\
  32 & 32 & -4 & -4 & 5 & -4 & 8 & 0 & 0 & 0 & -4 & 2 & 0 & -1 & 0 & 0 & 0 & 0 & 0 & 0 & 0 & $\dots$ \\
  $32'$ & 32 & -4 & -4 & 5 & -4 & 8 & 0 & 0 & 0 & -4 & 2 & 0 & -1 & 0 & 0 & 0 & 0 & 0 & 0 & 0 & $\dots$ \\
  $32''$ & 32 & -4 & -4 & 5 & -4 & 8 & 0 & 0 & 0 & -4 & 2 & 0 & -1 & 0 & 0 & 0 & 0 & 0 & 0 & 0 & $\dots$ \\
  $\widetilde{32}$ & 32 & -4 & -4 & 5 & -4 & -8 & 0 & 0 & 0 & 4 & -2 & 0 & 1 & 0 & 0 & 0 & 0 & 0 & 0 & 0 & $\dots$ \\
  $\widetilde{32}'$ & 32 & -4 & -4 & 5 & -4 & -8 & 0 & 0 & 0 & 4 & -2 & 0 & 1 & 0 & 0 & 0 & 0 & 0 & 0 & 0 & $\dots$ \\
  $\widetilde{32}''$ & 32 & -4 & -4 & 5 & -4 & -8 & 0 & 0 & 0 & 4 & -2 & 0 & 1 & 0 & 0 & 0 & 0 & 0 & 0 & 0 & $\dots$ \\
  48 & 48 & -24 & 12 & -6 & 3 & 0 & 0 & 0 & 0 & 0 & 0 & 0 & 0 & 0 & 0 & -4 & 2 & -1 & 0 & 0 & $\dots$ \\
  $\widetilde{48}$ & 48 & 12 & -6 & -6 & 12 & 0 & -8 & 0 & 0 & 0 & 0 & 4 & 0 & -2 & 0 & 0 & 0 & 0 & 0 & 0 & $\dots$ \\
  \hline
\end{tabular}}
{\tiny\begin{tabular}{|c|c|c|c|c|c|c|c|c|c|c|c|c|c|c|c|c|c|c|c|}
  \hline
  $~$  & 972 & 144 & 288 & 576 & 432 & 1296 & 864 & 288 & 432 & 864 & 144 & 288 & 576 & 432 & 1296 & 864 & 288 & 432 & 864
  \\\hline
  1 & 1 & 1 & 1 & 1 & 1 & 1 & 1 & 1 & 1 & 1 & 1 & 1 & 1 & 1 & 1 & 1 & 1 & 1 & 1 \\
  $1'$ & 1 & $\omega$ & $\omega$ & $\omega$ & $\omega$ & $\omega$ & $\omega$ & $\omega$ & $\omega$ & $\omega$ & $\omega^2$ & $\omega^2$ & $\omega^2$ & $\omega^2$ & $\omega^2$ & $\omega^2$ & $\omega^2$ & $\omega^2$ & $\omega^2$ \\
  $1''$ & 1 & $\omega^2$ & $\omega^2$ & $\omega^2$ & $\omega^2$ & $\omega^2$ & $\omega^2$ & $\omega^2$ & $\omega^2$ & $\omega^2$ & $\omega$ & $\omega$ & $\omega$ & $\omega$ & $\omega$ & $\omega$ & $\omega$ & $\omega$ & $\omega$ \\
  3 & -1 & 0 & 0 & 0 & 0 & 0 & 0 & 0 & 0 & 0 & 0 & 0 & 0 & 0 & 0 & 0 & 0 & 0 & 0 \\
  $\widetilde{1}$ & 1 & 1 & 1 & 1 & -1 & 1 & -1 & 1 & -1 & -1 & 1 & 1 & 1 & -1 & 1 & -1 & 1 & -1 & -1 \\
  $\widetilde{1}'$ & 1 & $\omega$ & $\omega$ & $\omega$ & -$\omega$ & $\omega$ & -$\omega$ & $\omega$ & -$\omega$ & -$\omega$ & $\omega^2$ & $\omega^2$ & $\omega^2$ & -$\omega^2$ & $\omega^2$ & -$\omega^2$ & $\omega^2$ & -$\omega^2$ & -$\omega^2$ \\
  $\widetilde{1}''$ & 1 & $\omega^2$ & $\omega^2$ & $\omega^2$ & -$\omega^2$ & $\omega^2$ & -$\omega^2$ & $\omega^2$ & -$\omega^2$ & -$\omega^2$ & $\omega$ & $\omega$ & $\omega$ & -$\omega$ & $\omega$ & -$\omega$ & $\omega$ & -$\omega$ & -$\omega$ \\
  $\widetilde{3}$ & -1 & 0 & 0 & 0 & 0 & 0 & 0 & 0 & 0 & 0 & 0 & 0 & 0 & 0 & 0 & 0 & 0 & 0 & 0 \\
  4 & 0 & 1 & 1 & 1 & -1 & -1 & 1 & 1 & 1 & -1 & 1 & 1 & 1 & -1 & -1 & 1 & 1 & 1 & -1 \\
  $4'$ & 0 & $\omega$ & $\omega$ & $\omega$ & -$\omega$ & -$\omega$ & $\omega$ & $\omega$ & $\omega$ & -$\omega$ & $\omega^2$ & $\omega^2$ & $\omega^2$ & -$\omega^2$ & -$\omega^2$ & $\omega^2$ & $\omega^2$ & $\omega^2$ & -$\omega^2$ \\
  $4''$ & 0 & $\omega^2$ & $\omega^2$ & $\omega^2$ & -$\omega^2$ & -$\omega^2$ & $\omega^2$ & $\omega^2$ & $\omega^2$ & -$\omega^2$ & $\omega$ & $\omega$ & $\omega$ & -$\omega$ & -$\omega$ & $\omega$ & $\omega$ & $\omega$ & -$\omega$ \\
  $\widetilde{4}$ & 0 & 1 & 1 & 1 & 1 & -1 & -1 & 1 & -1 & 1 & 1 & 1 & 1 & 1 & -1 & -1 & 1 & -1 & 1 \\
  $\widetilde{4}'$ & 0 & $\omega$ & $\omega$ & $\omega$ & $\omega$ & -$\omega$ & -$\omega$ & $\omega$ & -$\omega$ & $\omega$ & $\omega^2$ & $\omega^2$ & $\omega^2$ & $\omega^2$ & -$\omega^2$ & -$\omega^2$ & $\omega^2$ & -$\omega^2$ & $\omega^2$ \\
  $\widetilde{4}''$ & 0 & $\omega^2$ & $\omega^2$ & $\omega^2$ & $\omega^2$ & -$\omega^2$ & -$\omega^2$ & $\omega^2$ & -$\omega^2$ & $\omega^2$ & $\omega$ & $\omega$ & $\omega$ & $\omega$ & -$\omega$ & -$\omega$ & $\omega$ & -$\omega$ & $\omega$ \\
  6 & -2 & 0 & 0 & 0 & 0 & 0 & 0 & 0 & 0 & 0 & 0 & 0 & 0 & 0 & 0 & 0 & 0 & 0 & 0 \\
  $6'$ & 2 & 0 & 0 & 0 & 0 & 0 & 0 & 0 & 0 & 0 & 0 & 0 & 0 & 0 & 0 & 0 & 0 & 0 & 0 \\
  8 & 0 & 2 & -1 & -1 & 0 & 0 & -1 & 2 & 2 & 0 & 2 & -1 & -1 & 0 & 0 & -1 & 2 & 2 & 0 \\
  $8'$ & 0 & 2$\omega$ & -$\omega$ & -$\omega$ & 0 & 0 & -$\omega$ & 2$\omega$ & 2$\omega$ & 0 & 2$\omega^2$ & -$\omega^2$ & -$\omega^2$ & 0 & 0 & -$\omega^2$ & 2$\omega^2$ & 2$\omega^2$ & 0 \\
  $8''$ & 0 & 2$\omega^2$ & -$\omega^2$ & -$\omega^2$ & 0 & 0 & -$\omega^2$ & 2$\omega^2$ & 2$\omega^2$ & 0 & 2$\omega$ & -$\omega$ & -$\omega$ & 0 & 0 & -$\omega$ & 2$\omega$ & 2$\omega$ & 0 \\
  $\widetilde{8}$ & 0 & 2 & -1 & -1 & 0 & 0 & 1 & 2 & -2 & 0 & 2 & -1 & -1 & 0 & 0 & 1 & 2 & -2 & 0 \\
  $\widetilde{8}'$ & 0 & 2$\omega$ & -$\omega$ & -$\omega$ & 0 & 0 & $\omega$ & 2$\omega$ & -2$\omega$ & 0 & 2$\omega^2$ & -$\omega^2$ & -$\omega^2$ & 0 & 0 & $\omega^2$ & 2$\omega^2$ & -2$\omega^2$ & 0 \\
  $\widetilde{8}''$ & 0 & 2$\omega^2$ & -$\omega^2$ & -$\omega^2$ & 0 & 0 & $\omega^2$ & 2$\omega^2$ & -2$\omega^2$ & 0 & 2$\omega$ & -$\omega$ & -$\omega$ & 0 & 0 & $\omega$ & 2$\omega$ & -2$\omega$ & 0 \\
  16 & 0 & 4 & -2 & 1 & 0 & 0 & 0 & -2 & 0 & 0 & 4 & -2 & 1 & 0 & 0 & 0 & -2 & 0 & 0 \\
  $16'$ & 0 & 4$\omega$ & -2$\omega$ & $\omega$ & 0 & 0 & 0 & -2$\omega$ & 0 & 0 & 4$\omega^2$ & -2$\omega^2$ & $\omega^2$ & 0 & 0 & 0 & -2$\omega^2$ & 0 & 0 \\
  $16''$ & 0 & 4$\omega^2$ & -2$\omega^2$ & $\omega^2$ & 0 & 0 & 0 & -2$\omega^2$ & 0 & 0 & 4$\omega$ & -2$\omega$ & $\omega$ & 0 & 0 & 0 & -2$\omega$ & 0 & 0 \\
  24 & 0 & 0 & 0 & 0 & 0 & 0 & 0 & 0 & 0 & 0 & 0 & 0 & 0 & 0 & 0 & 0 & 0 & 0 & 0 \\
  $24'$ & 0 & 0 & 0 & 0 & 0 & 0 & 0 & 0 & 0 & 0 & 0 & 0 & 0 & 0 & 0 & 0 & 0 & 0 & 0 \\
  $\widetilde{24}$ & 0 & 0 & 0 & 0 & 0 & 0 & 0 & 0 & 0 & 0 & 0 & 0 & 0 & 0 & 0 & 0 & 0 & 0 & 0 \\
  $\widetilde{24}'$ & 0 & 0 & 0 & 0 & 0 & 0 & 0 & 0 & 0 & 0 & 0 & 0 & 0 & 0 & 0 & 0 & 0 & 0 & 0 \\
  $24''$ & 0 & 0 & 0 & 0 & 0 & 0 & 0 & 0 & 0 & 0 & 0 & 0 & 0 & 0 & 0 & 0 & 0 & 0 & 0  \\
  $24'''$ & 0 & 0 & 0 & 0 & 0 & 0 & 0 & 0 & 0 & 0 & 0 & 0 & 0 & 0 & 0 & 0 & 0 & 0 & 0 \\
  32 & 0 & 2 & 2 & -1 & 2 & 0 & 0 & -1 & 0 & -1 & 2 & 2 & -1 & 2 & 0 & 0 & -1 & 0 & -1 \\
  $32'$ & 0 & 2$\omega$ & 2$\omega$ & -$\omega$ & 2$\omega$ & 0 & 0 & -$\omega$ & 0 & -$\omega$ & 2$\omega^2$ & 2$\omega^2$ & -$\omega^2$ & 2$\omega^2$ & 0 & 0 & -$\omega^2$ & 0 & -$\omega^2$ \\
  $32''$ & 0 & 2$\omega^2$ & 2$\omega^2$ & -$\omega^2$ & 2$\omega^2$ & 0 & 0 & -$\omega^2$ & 0 & -$\omega^2$ & 2$\omega$ & 2$\omega$ & -$\omega$ & 2$\omega$ & 0 & 0 & -$\omega$ & 0 & -$\omega$ \\
  $\widetilde{32}$ & 0 & 2 & 2 & -1 & -2 & 0 & 0 & -1 & 0 & 1 & 2 & 2 & -1 & -2 & 0 & 0 & -1 & 0 & 1 \\
  $\widetilde{32}'$ & 0 & 2$\omega$ & 2$\omega$ & -$\omega$ & -2$\omega$ & 0 & 0 & -$\omega$ & 0 & $\omega$ & 2$\omega^2$ & 2$\omega^2$ & -$\omega^2$ & -2$\omega^2$ & 0 & 0 & -$\omega^2$ & 0 & $\omega^2$ \\
  $\widetilde{32}''$ & 0 & 2$\omega^2$ & 2$\omega^2$ & -$\omega^2$ & -2$\omega^2$ & 0 & 0 & -$\omega^2$ & 0 & $\omega^2$ & 2$\omega$ & 2$\omega$ & -$\omega$ & -2$\omega$ & 0 & 0 & -$\omega$ & 0 & $\omega$ \\
  48 & 0 & 0 & 0 & 0 & 0 & 0 & 0 & 0 & 0 & 0 & 0 & 0 & 0 & 0 & 0 & 0 & 0 & 0 & 0 \\
  $\widetilde{48}$ & 0 & 0 & 0 & 0 & 0 & 0 & 0 & 0 & 0 & 0 & 0 & 0 & 0 & 0 & 0 & 0 & 0 & 0 & 0 \\
  \hline
\end{tabular}}
\end{center}
\vspace*{-.15in}
\caption{The character table for
$(S_3\times S_3\times S_3\times S_3)\rtimes A_4$.}
\end{table}

\clearpage

\subsection{Lepton assignment}

\begin{table}[htb]
\begin{center}
\begin{tabular}{|c|c|c|c|c|}
  \hline
  $~$ & $SU(2)_L$ & $U(1)_Y$ & $S_3^4\rtimes A_4$ & $Z_2$ \\\hline
  $L$ & 2 & -1/2 & 3 & +1 \\
  $e_{R1} $ & 1 & -1 & 1 & -1 \\
  $e_{R2} $ & 1 & -1 & $1'$ & -1 \\
  $e_{R3} $ & 1 & -1 & $1''$ & -1 \\
  $N$ & 1 & 0 & 3 & +1 \\
  $N'$ & 1 & 0 & 48 & +1 \\
  $N''$ & 1 & 0 & 8 & +1 \\
  $E_L$ & 1 & -1 & 3 & +1 \\
  $E_R$ & 1 & -1 & 3 & +1 \\
  $\phi$ & 1 & 0 & 16 & +1 \\
  $\chi$ & 1 & 0 & 3 & -1 \\
  $H$ & 2 & -1/2 & 1 & +1 \\
  \hline
\end{tabular}
\end{center}
\caption{This table shows the assignments of the fermions and Higgs fields under
$SU(2)_L\times U(1)_Y\times(S_3^4\rtimes A_4)\times
Z_2$.}\end{table}

The leptonic and scalar particle content of the model in is given in Table 3.  In more detail, the SM leptons transform under $(S_3^4\rtimes A_4)\times Z_2$ as
\begin{eqnarray}e_{R1}\sim(1,-1),~~~~e_{R2}\sim(1',-1),~~~~e_{R3}\sim(1'',-1),~~~~(L_1,L_2,L_3)\sim
(3,+1).\end{eqnarray}
Here the second entry indicates $Z_2$ transformation of the field.
The $A_4$ assignment is identical to the one in
Ref. \cite{ma-rajasekaran,ma2002,babu-ma-valle,altarelli,babu-he,he-keum-volkas,ma2009,hmv,fhm,af2010}.
The charged lepton masses are generated by effective interactions involving a
real Higgs multiplet $\chi$ belonging to a $Z_2$ odd triplet of
$A_4$.  This involves integrating out multiplets of
heavy vector-like fermions, whose masses are at the high scale
$M_*$, with the same gauge quantum numbers as right-handed charged
leptons. These are $E_{L,R}\sim(3,+1)$ under $(S_3^4\rtimes
A_4)\times Z_2$. The Yukawa interactions for the charged leptons are
given by
\begin{eqnarray}
{\cal L}_{e}&=&\kappa
f_2(\overline{E}_{R},L)H+m_Ef_2(\overline{E}_{R},E_{L})+\epsilon_1\overline{e}_{R1}f_2(E_L,\chi)
\nonumber \\
&+& \epsilon_2g_5(\overline{e}_{R2},E_L,\chi) + \epsilon_3
g_6(\overline{e}_{R3},E_L,\chi)+c.c., \label{Le}
\end{eqnarray}
where the functions $f_2$, $g_5$, and $g_6$ are given in Appendix B.  Here $H$ is the SM Higgs doublet.

Neutrino Dirac masses are generated by effective interactions involving a real Higgs multiplet $\phi$
belonging to a $Z_2$ even $16$-dimensional representation of
$S_3^4\rtimes A_4$.
This involves integrating out
multiplets of heavy right-handed neutrinos, with masses at the high
scale $M_*$. These
multiplets are $N\sim(3,+1)$, $N'\sim(48,+1)$, and $N''\sim(8,+1)$
under $(S_3^4\rtimes A_4)\times Z_2$. The Yukawa interactions for
the neutrinos are given by
\begin{eqnarray}
{\cal L}_{\nu} &=&\lambda
f_2(L,N)\widetilde{H}+m_Nf_2(N,N)+m'_Nf_3(N',N')
+m''_Nf_4(N'',N'') \nonumber \\
&+& \alpha_1g_2(N,\phi,N')
+ \alpha_2g_3(N'',\phi,N')+\beta
g_4(\phi,N',N'),
\label{Lnu}
\end{eqnarray}
where the functions $f_2$, $f_3$, $f_4$, $g_2$, $g_3$, and $g_4$ are
given in Appendix B, and where $\widetilde{H} = i \tau_2 H^*$.
Interactions involving $\chi$, such as $NN\chi$ and $N'N'\chi$, are
allowed by $S_3^4\rtimes A_4$, but are prevented by the $Z_2$ symmetry.

As will be shown in the subsequent sections, these leptonic
interactions will lead to the desired tribimaximal pattern of
mixing, without any vacuum alignment problem. Symmetry-invariant
interactions between $\phi$ and $\chi$ must consist of products of
$S_3^4$ invariants constructed from $\phi$ with $Z_2$ invariants
constructed from $\chi$. The $16$-dimensional representation to
which $\phi$ belongs is $(2,2,2,2)$ with respect to $S_3^4$. (Since
$(2,2,2,2)$ is invariant under permutations, it does not mix with
other representations when $S_3^4$ is embedded into $S_3^4\rtimes
A_4$.) There is only one quadratic $S_3^4$ invariant that can be
constructed with $\phi$, which is a $1$ of $A_4$. So there is no
$16\times16\times3$ invariant. There is also only one cubic $S_3^4$
invariant that can be constructed with $\phi$, which is also a $1$
of $A_4$. So there is no $16\times16\times16\times3$ invariant.
Thus, there are no cubic invariants involving both $\phi$ and
$\chi$, and the only quartic invariant containing both is a product
of quadratic invariants, which does not generate a VEV alignment
problem. However, $16\times16\times16\times16$ does contain not only
$3$ but also $1'$ and $1''$, so that non-trivial $\phi^4\chi^2$
invariants do exist, but these are non-renormalizable interactions,
and are presumably suppressed by the Planck scale. Inclusion of such
suppressed operators will have no significant effects on the the
desired vacuum structure.

It is vitally important here that the $16$ contains only one
complete multiplet under the first factor, $S_3^4$, of the
semi-direct product. If an abelian group were used for the first
factor, then a representation that does not mix like this would have
to be one-dimensional. But it would be impossible to obtain the
symmetry breaking pattern we desire with a one-dimensional
representation. (A one-dimensional representation can only break a
group to an invariant subgroup, but the $Z_2$ and $Z_3$ subgroups of
$A_4$, referred to in the introduction, that we desire are not
invariant.) So, a non-abelian group is required, and $S_3$ is the
simplest non-abelian group. Four factors are used so that a
non-trivial semi-direct product with $A_4$ can be taken.

\section{Higgs potential and its minimum}

The potential involving
the $\phi$ and $\chi$ fields  has the form
\begin{eqnarray}V_{\phi\chi}&=&a_1f_1(\phi,\phi)+a_2f_2(\chi,\chi)+bg_1(\phi,\phi,\phi)
+c_1h_1(\phi)+c_2h_2(\phi)\nonumber \\
&+& c_3h_3(\chi)+c_4h_4(\chi)
+ c_5f_1(\phi,\phi)f_2(\chi,\chi),
\label{pot}
\end{eqnarray}
where the functions $f_1$, $f_2$, $g_1$, $h_1$, $h_2$, $h_3$, and
$h_4$ are given in Appendix B.

If the Higgs potential in (\ref{pot}) is to be bounded below, the
following constraints on the parameters must be satisfied:
\begin{eqnarray}c_1>0,~~~~c_1+2c_2>0,~~~~c_3>0,~~~~c_3+c_4>0,\nonumber \\
c_5+\sqrt{c_1c_3}>0,~~~~c_5+\frac{1}{2}\sqrt{\frac{1}{6}(c_1+2c_2)(c_3+c_4)}>0.
\label{bound}
\end{eqnarray}
Upon minimizing the potential,
\begin{eqnarray}\langle\chi\rangle &=& (v_{\chi},v_{\chi},v_{\chi}),\nonumber \\
\langle\phi\rangle &=& (0,0,0,0,v_{\phi},v_{\phi},v_{\phi},v_{\phi},
v_{\phi},v_{\phi},v_{\phi},v_{\phi},0,0,0,0)
\label{vev}
\end{eqnarray} is found
to be an extremum for
\begin{eqnarray}2a_1+3b_1v_{\phi}+2(c_1+c_2)v_{\phi}^2+6c_5v_{\chi}^2=0,
~~~~a_2+4(c_3+c_4)v_{\chi}^2+8c_5v_{\phi}^2=0.\end{eqnarray} The
Higgs boson masses are found to be
\begin{eqnarray}3bv_{\phi}+4(c_1+c_2)v_{\phi}^2+8(c_3+c_4)v_{\chi}^2\pm
\sqrt{[3bv_{\phi}+4(c_1+c_2)v_{\phi}^2-8(c_3+c_4)v_{\chi}^2]^2+1536c_5^2v_{\phi}^2v_{\chi}^2},\nonumber\end{eqnarray}
\begin{eqnarray}-18bv_{\phi},~~~-6bv_{\phi}-4c_1v_{\phi}^2,~~~6bv_{\phi}+8c_1v_{\phi}^2,
~~~6bv_{\phi}+8(c_1-c_2)v_{\phi}^2,~~~8(2c_3-c_4)v_{\chi}^2.\end{eqnarray}
Each mass in the first line occurs once, while those in the second
line occur four, eight, two, one, and two times respectively.
Assuming (\ref{bound}), we see that (\ref{vev}) is a local minimum for
\begin{eqnarray}&~& bv_{\phi}<0,~~~~\frac{2}{3}<\left|\frac{b}{c_1v_{\phi}}\right|<\frac{4}{3},
~~~~-1<\frac{c_3}{c_4}<2,~~~~-\frac{1}{3}<\frac{c_2/c_1}{4-3\left|\frac{b}{c_1v_{\phi}}\right|}<1,\nonumber\\
&~& |c_5|<\frac{1}{4}\sqrt{\frac{1}{3}\left[\frac{3b}{v_{\phi}}+4(c_1+c_2)\right](c_3+c_4)}\,\,.\end{eqnarray}

The VEV of $\chi$ in (\ref{vev}) breaks $A_4$ to the $Z_3$ subgroup of
$A_4$ generated by $Y$. Note also that $S_3^4$ is left unbroken by
$\chi$ trivially, but $Z_2$ is broken. On the other hand, the VEV of
$\phi$ breaks $A_4$ to the $Z_2$ subgroup of $A_4$ generated by $X$.
Additionally, $\phi$ leaves the generators $B_1$, $B_2$, $B_3B_4$,
and $A_3A_4$ unbroken. These form the subgroup $D_4\times S_3$, with
$D_4$ generated by $B_1$, $B_2$, and $X$ and with $S_3$ generated by
$A_3A_4$ and $B_3B_4$. This VEV also trivially leaves the additional
external $Z_2$ unbroken.

Note that the Higgs potential does not have any accidental global symmetries, which would
have resulted in pseudo-Goldstone bosons.  We conclude that consistent symmetry breaking can
be realized without any vacuum alignment problem.

\section{Lepton masses}

The light lepton masses can now be readily inferred from the Lagrangian
of (\ref{Le}) and (\ref{Lnu}), along with the VEV structure of $\chi$ and
$\phi$ given in (\ref{vev}).
The Higgs field $\chi$ mixes $E_{L,R}$ with the SM charged
leptons in (\ref{Le}), generating the light charged lepton mass matrix.
Upon integrating out $E_{L,R}$, the $Z_3$ subgroup of $A_4$ left
unbroken by the VEV of $\chi$ forces the light left-handed charged
lepton mass matrix to have the form (see Appendix C for details)
\begin{eqnarray}M_e^{\dag}M_e=\frac{1}{\sqrt{3}}\left(
                                     \begin{array}{ccc}
                                       1 & 1 & 1 \\
                                       1 & \omega & \omega^2 \\
                                       1 & \omega^2 & \omega \\
                                     \end{array}
                                   \right)\left(
                                            \begin{array}{ccc}
                                              m_e^2 & 0 & 0 \\
                                              0 & m_{\mu}^2 & 0 \\
                                              0 & 0 & m_{\tau}^2 \\
                                            \end{array}
                                          \right)\frac{1}{\sqrt{3}}\left(
                                     \begin{array}{ccc}
                                       1 & 1 & 1 \\
                                       1 & \omega^2 & \omega \\
                                       1 & \omega & \omega^2 \\
                                     \end{array}
                                   \right)\nonumber \end{eqnarray}
\begin{eqnarray}=U_L\left(
                       \begin{array}{ccc}
                       m_e^2 & 0 & 0 \\
                       0 & m_\mu^2 & 0 \\
                       0 & 0 & m_\tau^2 \\
                       \end{array}
                       \right)U_L^{\dag}.
\label{UL}
\end{eqnarray} The
masses are given by
\begin{eqnarray}m_i^2=\frac{3|\kappa\epsilon_iv_{\chi}v|^2}{3|\epsilon_iv_{\chi}|^2+|m_E|^2}.
\label{me}
\end{eqnarray}

The Higgs field $\phi$ mixes the heavy right-handed neutrinos with
the light left-handed neutrinos in (\ref{Lnu}), generating the light
neutrino mass matrix. $N\sim3$ is required because the SM Higgs $H$
only breaks EW symmetry, so that it can only cause left-handed
neutrinos to mix with a triplet. Since $3\times16=48$, $\phi\sim16$
induces mixing between $N$ and $N'\sim48$. $N''\sim8$ is needed to
remove unwanted accidental symmetries. Upon integrating out the
heavy right-handed neutrinos, the $Z_2$ subgroup of $A_4$ left
unbroken by the VEV of $\phi$ (along with an additional accidental
$Z_2$ that is actually part of $S_4$ \cite{A4vsS4}) forces the light
neutrino mass matrix to have the form (see Appendix D)
\begin{eqnarray}M_{\nu}=\left(
                          \begin{array}{ccc}
                            a_{\nu} & 0 & c_{\nu} \\
                            0 & b_{\nu} & 0 \\
                            c_{\nu} & 0 & a_{\nu} \\
                          \end{array}
                        \right).
\end{eqnarray}
This matrix is diagonalized by
\begin{eqnarray}U_{\nu}=\frac{1}{\sqrt{2}}\left(
                          \begin{array}{ccc}
                            1 & 0 & -1 \\
                            0 & \sqrt{2} & 0 \\
                            1 & 0 & 1 \\
                          \end{array}
                        \right)P_{\nu},
\label{Unu}
\end{eqnarray}
where diagonal $P_{\nu}$ is a phase matrix. The light neutrino
masses are found to be
\begin{eqnarray}m_1&=&\left|\frac{\lambda^2v^2}{2}~\frac{m'_N
m''_N-4\alpha_2^2v_{\phi}^2+\beta
v_{\phi}m''_N}{-2\alpha_1^2v_{\phi}^2+m_N \left(m'_N
m''_N-4\alpha_2^2v_{\phi}^2+\beta
v_{\phi}m''_N\right)}\right|,\nonumber \\
m_2&=&\left|\frac{\lambda^2v^2}{2}~\frac{m'_N
m''_N-2\alpha_2^2v_{\phi}^2+\beta
v_{\phi}m''_N}{-2\alpha_1^2v_{\phi}^2+m_N \left(m'_N
m''_N-2\alpha_2^2v_{\phi}^2+\beta
v_{\phi}m''_N\right)}\right|,\nonumber\\
m_3&=&\left|\frac{\lambda^2v^2}{2}~\frac{m'_N+\beta v_{\phi}}{-2\alpha_1^2v_{\phi}^2
+m_N (m'_N+\beta v_{\phi})}\right|.
\label{mnu}
\end{eqnarray}

Note that the three neutrino masses given in (\ref{mnu}) are unrelated.  This spectrum allows for
normal neutrino mass hierarchy, inverted mass hierarchy,
as well as quasi-degenerate neutrinos.  The last possibility is realized when the bare
mass parameters in (\ref{mnu}) are taken to be slightly larger than the VEV $v_\phi$.

Eqs. (\ref{UL}) and (\ref{Unu}) give the desired form (\ref{tribi}) for the mixing
matrix $U_{MNS}=U_L^TU_{\nu}^*$.

\section{Inclusion of quarks}

We have not yet specified how quarks transform under the finite family
symmetry. One possibility is to have the quarks not transform under
this symmetry at all, in which case the quark sector is exactly like
the SM. Another, perhaps more interesting, possibility is to use an assignment analogous to
that of the charged leptons,
\begin{eqnarray}u_{R1},d_{R1}\sim(1,-1),~~~u_{R2},d_{R2}\sim(1',-1),~~~u_{R3},d_{R3}\sim(1'',-1),~~~(Q_1,Q_2,Q_3)_L\sim
(3,+1).\nonumber\end{eqnarray} Then, add additional heavy quark
multiplets $U_{L,R}$ and $D_{L,R}$, transforming as
$(3,1,+2/3,3,+1)$ and $(3,1,-1/3,3,+1)$ respectively under
$SU(3)_c\times SU(2)_L\times U(1)_Y\times(S_3^4\rtimes A_4)\times
Z_2$. For the up-type quarks, we obtain the Yukawa Lagrangian
\begin{eqnarray}{\cal
L}_{u}&=&\kappa'
f_2(\overline{U}_{R},Q_L)H+m_Uf_2(\overline{U}_{R},U_{L})+\epsilon'_1\overline{u}_{R1}f_2(U_L,\chi)
\nonumber \\
&+&\epsilon'_2g_5(\overline{u}_{R2},U_L,\chi)
+\epsilon'_3g_6(\overline{u}_{R3},U_L,\chi)+c.c.,\nonumber\end{eqnarray}
with a similar result for down-type quarks. This gives
non-degenerate quark masses but no mixing.

Small and arbitrary quark mixing can be generated
from $\phi$ by adding additional heavy quark multiplets:
\begin{eqnarray}&~& U'_{1L,R}\sim(3,1,+2/3,16,-1),~~~U'_{2L,R}\sim(3,1,+2/3,16',-1),\nonumber \\
&~&U'_{3L,R}\sim(3,1,+2/3,16'',-1),
~~~U''_{L,R}\sim(3,1,+2/3,6,-1).\nonumber\end{eqnarray}
The generators for these representations can be found in Appendix A.
Heavy quarks can be added to the down-type sector instead, but to
generate mixing it is only necessary to do this for one sector. The
up-type Yukawa Lagrangian receives the additional contribution
\begin{eqnarray}{\cal
L}'_{u}&=&m'_{1U}f_1(\overline{U}'_{1R},U'_{1L})+m'_{2U}f_1(\overline{U}'_{2R},U'_{2L})
+m'_{3U}f_1(\overline{U}'_{3R},U'_{3L})+m''_{U}f_5(\overline{U}''_{R},U''_{L})\nonumber \\
&+&\delta_1\overline{u}_{R1}f_2(U'_L,\phi)
+\delta_2\overline{u}_{R2}f_2(U'_L,\phi)+\delta_3\overline{u}_{R3}f_2(U'_L,\phi)
+\zeta_1 g_7(\overline{U}''_{R},U'_{1L},\phi)+\zeta_2
g_8(\overline{U}''_{R},U'_{2L},\phi)\nonumber \\
&+&\zeta_3
g_9(\overline{U}''_{R},U'_{3L},\phi)+\xi_1
g_7(\overline{U}''_{L},U'_{1R},\phi)+\xi_2
g_8(\overline{U}''_{L},U'_{2R},\phi)+\xi_3
g_9(\overline{U}''_{L},U'_{3R},\phi),\nonumber\end{eqnarray} where
the functions $f_5$, $g_7$, $g_8$, and $g_9$ can be found in
Appendix B. In this way, arbitrary quark mixing can be obtained.

\section{Supersymmetrization}

Supersymmetrization of the $(S_3)^4\rtimes A_4$ sector of the model is straightforward.
Since in a renormaliable SUSY model one uses only quadratic and cubic
(and not quartic) invariants in
the superpotential, care should be given to ensure that there are no flat
directions.  The $Z_2$ factor in the model of Sec. II
would prevent a $\chi^3$ invariant. This
will leads to flat directions in the cubic superpotential of the
SUSY version of the model. If one wishes to avoid increasing the
number of Higgs multiplets, the problem can be solved by replacing
$Z_2$ with some other symmetry. So, consider $(S_3^4\rtimes
A_4)\times H$. It's not difficult to see that for $H=Z_n$, the flat
direction problem is not solved. The next simplest possibility is to
consider $\chi\sim(3,2)$ of $(S_3^4\rtimes A_4)\times H$, with $H$ a
$D_n$ or $Q_n$. However, one finds that the flat direction problem
persists for representations such as these. So, we are led to
consider $(S_3^4\rtimes A_4)\times A_4$ with $\chi\sim(3,3)$. It is
easily confirmed that the $(3,3)$ cubic superpotential has no flat
directions. The light charged lepton masses can be generated from this $\chi$ by
integrating out two sets of heavy vector-like superfields with the
same gauge quantum numbers as right-handed charged leptons,
transforming as $(3,1)$ and $(3,3)$ under $(S_3^4\rtimes A_4)\times
A_4$. Alternatively, one can solve the flat direction problem by
increasing the number on Higgs multiplets.

The SUSY flavor problem can be readily solved by the
symmetry $G= (S_3^4\rtimes A_4)\times A_4$.  The SM lepton and
quark doublets belong to triplet representations of the $A_4$ factor
group, which would mean that the soft SUSY breaking masses of all three
families of sleptons (and similarly squarks) are the same.  As for the
right--handed sleptons and squarks, the group $G$ does allow the three
families to have non-degenerate soft masses.  However, there is no
mixing in the right--handed slepton mass matrix, which is sufficient
to ensure the absence of excessive SUSY flavor violation.  Since there is mixing
between the right--handed singlet leptons and heavy vector-like leptons
of order $v_{\chi,\phi}/M_*$, adequate suppression of SUSY FCNC would require
that $v_{\chi,\phi}$ be somewhat smaller than $M_*$.

\section{Summary and conclusions}

First, we presented a renormalizable non-supersymmetric model based
on the finite symmetry $(S_3^4\rtimes A_4)\times Z_2$, with the SM
leptons assigned to representations of $A_4$. Neutrino masses are
generated by a Higgs field $\phi$ belonging to a $Z_2$ even
$16$-dimensional representation of $S_3^4\rtimes A_4$ while
charged-lepton masses are generated by a Higgs field $\chi$
belonging to a $Z_2$ odd triplet representation of $A_4$. The
additional symmetries, $S_3^4$ and $Z_2$, prevent quadratic and
cubic interactions between $\phi$ and $\chi$ and allow only a
trivial quartic interaction that does not cause an alignment
problem, addressing the alignment problem without altering the
desired properties of the family symmetry. In this way, we are able
to explain all aspects of neutrino mixing using only symmetries
which are spontaneously broken by the Higgs mechanism.

\section*{Appendix A}

In this appendix, we give the explicit matrices representing the
generators in Eqs. (\ref{group1})-(\ref{group4}) for each representation used in our
model.  These matrices will be used for computing the group invariants, given in Appendix B.
The generators for $16$, in a certain basis, are
{\small\begin{eqnarray}A_1^{(16)}=diag(\omega,\omega^2,\omega,\omega^2,\omega,\omega^2,\omega,\omega^2,
\omega,\omega^2,\omega,\omega^2,\omega,\omega^2,\omega,\omega^2),\nonumber\end{eqnarray}}
\vspace{-0.5cm}
{\small\begin{eqnarray}A_2^{(16)}=diag(\omega,\omega,\omega^2,\omega^2,\omega,\omega,\omega^2,\omega^2,
\omega,\omega,\omega^2,\omega^2,\omega,\omega,\omega^2,\omega^2),\nonumber\end{eqnarray}}
\vspace{-0.5cm}
{\small\begin{eqnarray}A_3^{(16)}=diag(\omega,\omega,\omega,\omega,\omega^2,\omega^2,\omega^2,\omega^2,
\omega,\omega,\omega,\omega,\omega^2,\omega^2,\omega^2,\omega^2),\nonumber\end{eqnarray}}
\vspace{-0.5cm}
{\small\begin{eqnarray}A_4^{(16)}=diag(\omega,\omega,\omega,\omega,\omega,\omega,\omega,\omega,
\omega^2,\omega^2,\omega^2,\omega^2,\omega^2,\omega^2,\omega^2,\omega^2),\nonumber\end{eqnarray}}
\vspace{-0.5cm} {\small\begin{eqnarray}B_1^{(16)}=\left(
\begin{array}{cccccccccccccccc}
 0 & 1 & 0 & 0 & 0 & 0 & 0 & 0 & 0 & 0 & 0 & 0 & 0 & 0 & 0 & 0 \\
 1 & 0 & 0 & 0 & 0 & 0 & 0 & 0 & 0 & 0 & 0 & 0 & 0 & 0 & 0 & 0 \\
 0 & 0 & 0 & 1 & 0 & 0 & 0 & 0 & 0 & 0 & 0 & 0 & 0 & 0 & 0 & 0 \\
 0 & 0 & 1 & 0 & 0 & 0 & 0 & 0 & 0 & 0 & 0 & 0 & 0 & 0 & 0 & 0 \\
 0 & 0 & 0 & 0 & 0 & 1 & 0 & 0 & 0 & 0 & 0 & 0 & 0 & 0 & 0 & 0 \\
 0 & 0 & 0 & 0 & 1 & 0 & 0 & 0 & 0 & 0 & 0 & 0 & 0 & 0 & 0 & 0 \\
 0 & 0 & 0 & 0 & 0 & 0 & 0 & 1 & 0 & 0 & 0 & 0 & 0 & 0 & 0 & 0 \\
 0 & 0 & 0 & 0 & 0 & 0 & 1 & 0 & 0 & 0 & 0 & 0 & 0 & 0 & 0 & 0 \\
 0 & 0 & 0 & 0 & 0 & 0 & 0 & 0 & 0 & 1 & 0 & 0 & 0 & 0 & 0 & 0 \\
 0 & 0 & 0 & 0 & 0 & 0 & 0 & 0 & 1 & 0 & 0 & 0 & 0 & 0 & 0 & 0 \\
 0 & 0 & 0 & 0 & 0 & 0 & 0 & 0 & 0 & 0 & 0 & 1 & 0 & 0 & 0 & 0 \\
 0 & 0 & 0 & 0 & 0 & 0 & 0 & 0 & 0 & 0 & 1 & 0 & 0 & 0 & 0 & 0 \\
 0 & 0 & 0 & 0 & 0 & 0 & 0 & 0 & 0 & 0 & 0 & 0 & 0 & 1 & 0 & 0 \\
 0 & 0 & 0 & 0 & 0 & 0 & 0 & 0 & 0 & 0 & 0 & 0 & 1 & 0 & 0 & 0 \\
 0 & 0 & 0 & 0 & 0 & 0 & 0 & 0 & 0 & 0 & 0 & 0 & 0 & 0 & 0 & 1 \\
 0 & 0 & 0 & 0 & 0 & 0 & 0 & 0 & 0 & 0 & 0 & 0 & 0 & 0 & 1 & 0
\end{array}
\right),\nonumber\end{eqnarray}}
{\small\begin{eqnarray}B_2^{(16)}=\left(
\begin{array}{cccccccccccccccc}
 0 & 0 & 1 & 0 & 0 & 0 & 0 & 0 & 0 & 0 & 0 & 0 & 0 & 0 & 0 & 0 \\
 0 & 0 & 0 & 1 & 0 & 0 & 0 & 0 & 0 & 0 & 0 & 0 & 0 & 0 & 0 & 0 \\
 1 & 0 & 0 & 0 & 0 & 0 & 0 & 0 & 0 & 0 & 0 & 0 & 0 & 0 & 0 & 0 \\
 0 & 1 & 0 & 0 & 0 & 0 & 0 & 0 & 0 & 0 & 0 & 0 & 0 & 0 & 0 & 0 \\
 0 & 0 & 0 & 0 & 0 & 0 & 1 & 0 & 0 & 0 & 0 & 0 & 0 & 0 & 0 & 0 \\
 0 & 0 & 0 & 0 & 0 & 0 & 0 & 1 & 0 & 0 & 0 & 0 & 0 & 0 & 0 & 0 \\
 0 & 0 & 0 & 0 & 1 & 0 & 0 & 0 & 0 & 0 & 0 & 0 & 0 & 0 & 0 & 0 \\
 0 & 0 & 0 & 0 & 0 & 1 & 0 & 0 & 0 & 0 & 0 & 0 & 0 & 0 & 0 & 0 \\
 0 & 0 & 0 & 0 & 0 & 0 & 0 & 0 & 0 & 0 & 1 & 0 & 0 & 0 & 0 & 0 \\
 0 & 0 & 0 & 0 & 0 & 0 & 0 & 0 & 0 & 0 & 0 & 1 & 0 & 0 & 0 & 0 \\
 0 & 0 & 0 & 0 & 0 & 0 & 0 & 0 & 1 & 0 & 0 & 0 & 0 & 0 & 0 & 0 \\
 0 & 0 & 0 & 0 & 0 & 0 & 0 & 0 & 0 & 1 & 0 & 0 & 0 & 0 & 0 & 0 \\
 0 & 0 & 0 & 0 & 0 & 0 & 0 & 0 & 0 & 0 & 0 & 0 & 0 & 0 & 1 & 0 \\
 0 & 0 & 0 & 0 & 0 & 0 & 0 & 0 & 0 & 0 & 0 & 0 & 0 & 0 & 0 & 1 \\
 0 & 0 & 0 & 0 & 0 & 0 & 0 & 0 & 0 & 0 & 0 & 0 & 1 & 0 & 0 & 0 \\
 0 & 0 & 0 & 0 & 0 & 0 & 0 & 0 & 0 & 0 & 0 & 0 & 0 & 1 & 0 & 0
\end{array}
\right),\nonumber\end{eqnarray}}
{\small\begin{eqnarray}B_3^{(16)}=\left(
\begin{array}{cccccccccccccccc}
 0 & 0 & 0 & 0 & 1 & 0 & 0 & 0 & 0 & 0 & 0 & 0 & 0 & 0 & 0 & 0 \\
 0 & 0 & 0 & 0 & 0 & 1 & 0 & 0 & 0 & 0 & 0 & 0 & 0 & 0 & 0 & 0 \\
 0 & 0 & 0 & 0 & 0 & 0 & 1 & 0 & 0 & 0 & 0 & 0 & 0 & 0 & 0 & 0 \\
 0 & 0 & 0 & 0 & 0 & 0 & 0 & 1 & 0 & 0 & 0 & 0 & 0 & 0 & 0 & 0 \\
 1 & 0 & 0 & 0 & 0 & 0 & 0 & 0 & 0 & 0 & 0 & 0 & 0 & 0 & 0 & 0 \\
 0 & 1 & 0 & 0 & 0 & 0 & 0 & 0 & 0 & 0 & 0 & 0 & 0 & 0 & 0 & 0 \\
 0 & 0 & 1 & 0 & 0 & 0 & 0 & 0 & 0 & 0 & 0 & 0 & 0 & 0 & 0 & 0 \\
 0 & 0 & 0 & 1 & 0 & 0 & 0 & 0 & 0 & 0 & 0 & 0 & 0 & 0 & 0 & 0 \\
 0 & 0 & 0 & 0 & 0 & 0 & 0 & 0 & 0 & 0 & 0 & 0 & 1 & 0 & 0 & 0 \\
 0 & 0 & 0 & 0 & 0 & 0 & 0 & 0 & 0 & 0 & 0 & 0 & 0 & 1 & 0 & 0 \\
 0 & 0 & 0 & 0 & 0 & 0 & 0 & 0 & 0 & 0 & 0 & 0 & 0 & 0 & 1 & 0 \\
 0 & 0 & 0 & 0 & 0 & 0 & 0 & 0 & 0 & 0 & 0 & 0 & 0 & 0 & 0 & 1 \\
 0 & 0 & 0 & 0 & 0 & 0 & 0 & 0 & 1 & 0 & 0 & 0 & 0 & 0 & 0 & 0 \\
 0 & 0 & 0 & 0 & 0 & 0 & 0 & 0 & 0 & 1 & 0 & 0 & 0 & 0 & 0 & 0 \\
 0 & 0 & 0 & 0 & 0 & 0 & 0 & 0 & 0 & 0 & 1 & 0 & 0 & 0 & 0 & 0 \\
 0 & 0 & 0 & 0 & 0 & 0 & 0 & 0 & 0 & 0 & 0 & 1 & 0 & 0 & 0 & 0
\end{array}
\right),\nonumber\end{eqnarray}}
{\small\begin{eqnarray}B_4^{(16)}=\left(
\begin{array}{cccccccccccccccc}
 0 & 0 & 0 & 0 & 0 & 0 & 0 & 0 & 1 & 0 & 0 & 0 & 0 & 0 & 0 & 0 \\
 0 & 0 & 0 & 0 & 0 & 0 & 0 & 0 & 0 & 1 & 0 & 0 & 0 & 0 & 0 & 0 \\
 0 & 0 & 0 & 0 & 0 & 0 & 0 & 0 & 0 & 0 & 1 & 0 & 0 & 0 & 0 & 0 \\
 0 & 0 & 0 & 0 & 0 & 0 & 0 & 0 & 0 & 0 & 0 & 1 & 0 & 0 & 0 & 0 \\
 0 & 0 & 0 & 0 & 0 & 0 & 0 & 0 & 0 & 0 & 0 & 0 & 1 & 0 & 0 & 0 \\
 0 & 0 & 0 & 0 & 0 & 0 & 0 & 0 & 0 & 0 & 0 & 0 & 0 & 1 & 0 & 0 \\
 0 & 0 & 0 & 0 & 0 & 0 & 0 & 0 & 0 & 0 & 0 & 0 & 0 & 0 & 1 & 0 \\
 0 & 0 & 0 & 0 & 0 & 0 & 0 & 0 & 0 & 0 & 0 & 0 & 0 & 0 & 0 & 1 \\
 1 & 0 & 0 & 0 & 0 & 0 & 0 & 0 & 0 & 0 & 0 & 0 & 0 & 0 & 0 & 0 \\
 0 & 1 & 0 & 0 & 0 & 0 & 0 & 0 & 0 & 0 & 0 & 0 & 0 & 0 & 0 & 0 \\
 0 & 0 & 1 & 0 & 0 & 0 & 0 & 0 & 0 & 0 & 0 & 0 & 0 & 0 & 0 & 0 \\
 0 & 0 & 0 & 1 & 0 & 0 & 0 & 0 & 0 & 0 & 0 & 0 & 0 & 0 & 0 & 0 \\
 0 & 0 & 0 & 0 & 1 & 0 & 0 & 0 & 0 & 0 & 0 & 0 & 0 & 0 & 0 & 0 \\
 0 & 0 & 0 & 0 & 0 & 1 & 0 & 0 & 0 & 0 & 0 & 0 & 0 & 0 & 0 & 0 \\
 0 & 0 & 0 & 0 & 0 & 0 & 1 & 0 & 0 & 0 & 0 & 0 & 0 & 0 & 0 & 0 \\
 0 & 0 & 0 & 0 & 0 & 0 & 0 & 1 & 0 & 0 & 0 & 0 & 0 & 0 & 0 & 0
\end{array}
\right),\nonumber\end{eqnarray}}
{\small\begin{eqnarray}X^{(16)}=\left(
\begin{array}{cccccccccccccccc}
 1 & 0 & 0 & 0 & 0 & 0 & 0 & 0 & 0 & 0 & 0 & 0 & 0 & 0 & 0 & 0 \\
 0 & 0 & 1 & 0 & 0 & 0 & 0 & 0 & 0 & 0 & 0 & 0 & 0 & 0 & 0 & 0 \\
 0 & 1 & 0 & 0 & 0 & 0 & 0 & 0 & 0 & 0 & 0 & 0 & 0 & 0 & 0 & 0 \\
 0 & 0 & 0 & 1 & 0 & 0 & 0 & 0 & 0 & 0 & 0 & 0 & 0 & 0 & 0 & 0 \\
 0 & 0 & 0 & 0 & 0 & 0 & 0 & 0 & 1 & 0 & 0 & 0 & 0 & 0 & 0 & 0 \\
 0 & 0 & 0 & 0 & 0 & 0 & 0 & 0 & 0 & 0 & 1 & 0 & 0 & 0 & 0 & 0 \\
 0 & 0 & 0 & 0 & 0 & 0 & 0 & 0 & 0 & 1 & 0 & 0 & 0 & 0 & 0 & 0 \\
 0 & 0 & 0 & 0 & 0 & 0 & 0 & 0 & 0 & 0 & 0 & 1 & 0 & 0 & 0 & 0 \\
 0 & 0 & 0 & 0 & 1 & 0 & 0 & 0 & 0 & 0 & 0 & 0 & 0 & 0 & 0 & 0 \\
 0 & 0 & 0 & 0 & 0 & 0 & 1 & 0 & 0 & 0 & 0 & 0 & 0 & 0 & 0 & 0 \\
 0 & 0 & 0 & 0 & 0 & 1 & 0 & 0 & 0 & 0 & 0 & 0 & 0 & 0 & 0 & 0 \\
 0 & 0 & 0 & 0 & 0 & 0 & 0 & 1 & 0 & 0 & 0 & 0 & 0 & 0 & 0 & 0 \\
 0 & 0 & 0 & 0 & 0 & 0 & 0 & 0 & 0 & 0 & 0 & 0 & 1 & 0 & 0 & 0 \\
 0 & 0 & 0 & 0 & 0 & 0 & 0 & 0 & 0 & 0 & 0 & 0 & 0 & 0 & 1 & 0 \\
 0 & 0 & 0 & 0 & 0 & 0 & 0 & 0 & 0 & 0 & 0 & 0 & 0 & 1 & 0 & 0 \\
 0 & 0 & 0 & 0 & 0 & 0 & 0 & 0 & 0 & 0 & 0 & 0 & 0 & 0 & 0 & 1
\end{array}
\right),\nonumber\end{eqnarray}} \vspace{-0.5cm}
{\small\begin{eqnarray}Y^{(16)}=\left(
\begin{array}{cccccccccccccccc}
 1 & 0 & 0 & 0 & 0 & 0 & 0 & 0 & 0 & 0 & 0 & 0 & 0 & 0 & 0 & 0 \\
 0 & 1 & 0 & 0 & 0 & 0 & 0 & 0 & 0 & 0 & 0 & 0 & 0 & 0 & 0 & 0 \\
 0 & 0 & 0 & 0 & 0 & 0 & 0 & 0 & 1 & 0 & 0 & 0 & 0 & 0 & 0 & 0 \\
 0 & 0 & 0 & 0 & 0 & 0 & 0 & 0 & 0 & 1 & 0 & 0 & 0 & 0 & 0 & 0 \\
 0 & 0 & 1 & 0 & 0 & 0 & 0 & 0 & 0 & 0 & 0 & 0 & 0 & 0 & 0 & 0 \\
 0 & 0 & 0 & 1 & 0 & 0 & 0 & 0 & 0 & 0 & 0 & 0 & 0 & 0 & 0 & 0 \\
 0 & 0 & 0 & 0 & 0 & 0 & 0 & 0 & 0 & 0 & 1 & 0 & 0 & 0 & 0 & 0 \\
 0 & 0 & 0 & 0 & 0 & 0 & 0 & 0 & 0 & 0 & 0 & 1 & 0 & 0 & 0 & 0 \\
 0 & 0 & 0 & 0 & 1 & 0 & 0 & 0 & 0 & 0 & 0 & 0 & 0 & 0 & 0 & 0 \\
 0 & 0 & 0 & 0 & 0 & 1 & 0 & 0 & 0 & 0 & 0 & 0 & 0 & 0 & 0 & 0 \\
 0 & 0 & 0 & 0 & 0 & 0 & 0 & 0 & 0 & 0 & 0 & 0 & 1 & 0 & 0 & 0 \\
 0 & 0 & 0 & 0 & 0 & 0 & 0 & 0 & 0 & 0 & 0 & 0 & 0 & 1 & 0 & 0 \\
 0 & 0 & 0 & 0 & 0 & 0 & 1 & 0 & 0 & 0 & 0 & 0 & 0 & 0 & 0 & 0 \\
 0 & 0 & 0 & 0 & 0 & 0 & 0 & 1 & 0 & 0 & 0 & 0 & 0 & 0 & 0 & 0 \\
 0 & 0 & 0 & 0 & 0 & 0 & 0 & 0 & 0 & 0 & 0 & 0 & 0 & 0 & 1 & 0 \\
 0 & 0 & 0 & 0 & 0 & 0 & 0 & 0 & 0 & 0 & 0 & 0 & 0 & 0 & 0 & 1
\end{array}
\right).\nonumber\end{eqnarray}}The generators for $16'$ and $16''$
are the same except $Y^{(16')}=\omega Y^{(16)}$ and
$Y^{(16'')}=\omega^2Y^{(16)}$. The generators for $48$ are
\begin{eqnarray}A_1^{(48)}=\left(
                             \begin{array}{ccc}
                               A_1^{(16)} & 0 & 0 \\
                               0 & A_1^{(16)} & 0 \\
                               0 & 0 & A_1^{(16)} \\
                             \end{array}
                           \right),~~~~B_1^{(48)}=\left(
                             \begin{array}{ccc}
                               B_1^{(16)} & 0 & 0 \\
                               0 & B_1^{(16)} & 0 \\
                               0 & 0 & B_1^{(16)} \\
                             \end{array}
                           \right),
\nonumber\end{eqnarray}

\begin{eqnarray}A_2^{(48)}=\left(
                             \begin{array}{ccc}
                               A_2^{(16)} & 0 & 0 \\
                               0 & A_4^{(16)} & 0 \\
                               0 & 0 & A_3^{(16)} \\
                             \end{array}
                           \right),~~~~B_2^{(48)}=\left(
                             \begin{array}{ccc}
                               B_2^{(16)} & 0 & 0 \\
                               0 & B_4^{(16)} & 0 \\
                               0 & 0 & B_3^{(16)} \\
                             \end{array}
                           \right),
\nonumber\end{eqnarray}

\begin{eqnarray}A_3^{(48)}=\left(
                             \begin{array}{ccc}
                               A_3^{(16)} & 0 & 0 \\
                               0 & A_2^{(16)} & 0 \\
                               0 & 0 & A_4^{(16)} \\
                             \end{array}
                           \right),~~~~B_3^{(48)}=\left(
                             \begin{array}{ccc}
                               B_3^{(16)} & 0 & 0 \\
                               0 & B_2^{(16)} & 0 \\
                               0 & 0 & B_4^{(16)} \\
                             \end{array}
                           \right),
\nonumber\end{eqnarray}

\begin{eqnarray}A_4^{(48)}=\left(
                             \begin{array}{ccc}
                               A_4^{(16)} & 0 & 0 \\
                               0 & A_3^{(16)} & 0 \\
                               0 & 0 & A_2^{(16)} \\
                             \end{array}
                           \right),~~~~B_4^{(48)}=\left(
                             \begin{array}{ccc}
                               B_4^{(16)} & 0 & 0 \\
                               0 & B_3^{(16)} & 0 \\
                               0 & 0 & B_2^{(16)} \\
                             \end{array}
                           \right),
\nonumber\end{eqnarray}

\begin{eqnarray}X^{(48)}=\left(
                             \begin{array}{ccc}
                               X^{(16)} & 0 & 0 \\
                               0 & -X^{(16)}Y^{(16)}X^{(16)}(Y^{(16)})^2 & 0 \\
                               0 & 0 & -Y^{(16)}X^{(16)}(Y^{(16)})^2 \\
                             \end{array}
                           \right),~~Y^{(48)}=\left(
                             \begin{array}{ccc}
                               0 & 0 & I \\
                               I & 0 & 0 \\
                               0 & I & 0 \\
                             \end{array}
                           \right).
\nonumber\end{eqnarray} And the generators for $8$ are
\begin{eqnarray}A_1^{(8)}=diag(\omega,\omega^2,1,1,1,1,1,1),
~~~~A_2^{(8)}=diag(1,1,\omega,\omega^2,1,1,1,1),\nonumber\end{eqnarray}

\begin{eqnarray}A_3^{(8)}=diag(1,1,1,1,\omega,\omega^2,1,1),
~~~~A_4^{(8)}=diag(1,1,1,1,1,1,\omega,\omega^2),\nonumber\end{eqnarray}

\begin{eqnarray}B_1^{(8)}=\left(
\begin{array}{cccccccc}
 0 & 1 & 0 & 0 & 0 & 0 & 0 & 0 \\
 1 & 0 & 0 & 0 & 0 & 0 & 0 & 0 \\
 0 & 0 & 1 & 0 & 0 & 0 & 0 & 0 \\
 0 & 0 & 0 & 1 & 0 & 0 & 0 & 0 \\
 0 & 0 & 0 & 0 & 1 & 0 & 0 & 0 \\
 0 & 0 & 0 & 0 & 0 & 1 & 0 & 0 \\
 0 & 0 & 0 & 0 & 0 & 0 & 1 & 0 \\
 0 & 0 & 0 & 0 & 0 & 0 & 0 & 1
\end{array}
\right),~~~~B_2^{(8)}=\left(
\begin{array}{cccccccc}
 1 & 0 & 0 & 0 & 0 & 0 & 0 & 0 \\
 0 & 1 & 0 & 0 & 0 & 0 & 0 & 0 \\
 0 & 0 & 0 & 1 & 0 & 0 & 0 & 0 \\
 0 & 0 & 1 & 0 & 0 & 0 & 0 & 0 \\
 0 & 0 & 0 & 0 & 1 & 0 & 0 & 0 \\
 0 & 0 & 0 & 0 & 0 & 1 & 0 & 0 \\
 0 & 0 & 0 & 0 & 0 & 0 & 1 & 0 \\
 0 & 0 & 0 & 0 & 0 & 0 & 0 & 1
\end{array}
\right),\nonumber\end{eqnarray}

\begin{eqnarray}B_3^{(8)}=\left(
\begin{array}{cccccccc}
 1 & 0 & 0 & 0 & 0 & 0 & 0 & 0 \\
 0 & 1 & 0 & 0 & 0 & 0 & 0 & 0 \\
 0 & 0 & 1 & 0 & 0 & 0 & 0 & 0 \\
 0 & 0 & 0 & 1 & 0 & 0 & 0 & 0 \\
 0 & 0 & 0 & 0 & 0 & 1 & 0 & 0 \\
 0 & 0 & 0 & 0 & 1 & 0 & 0 & 0 \\
 0 & 0 & 0 & 0 & 0 & 0 & 1 & 0 \\
 0 & 0 & 0 & 0 & 0 & 0 & 0 & 1
\end{array}
\right),~~~~B_4^{(8)}=\left(
\begin{array}{cccccccc}
 1 & 0 & 0 & 0 & 0 & 0 & 0 & 0 \\
 0 & 1 & 0 & 0 & 0 & 0 & 0 & 0 \\
 0 & 0 & 1 & 0 & 0 & 0 & 0 & 0 \\
 0 & 0 & 0 & 1 & 0 & 0 & 0 & 0 \\
 0 & 0 & 0 & 0 & 1 & 0 & 0 & 0 \\
 0 & 0 & 0 & 0 & 0 & 1 & 0 & 0 \\
 0 & 0 & 0 & 0 & 0 & 0 & 0 & 1 \\
 0 & 0 & 0 & 0 & 0 & 0 & 1 & 0
\end{array}
\right),\nonumber\end{eqnarray}

\begin{eqnarray}X^{(8)}=\left(
\begin{array}{cccccccc}
 0 & 0 & 1 & 0 & 0 & 0 & 0 & 0 \\
 0 & 0 & 0 & 1 & 0 & 0 & 0 & 0 \\
 1 & 0 & 0 & 0 & 0 & 0 & 0 & 0 \\
 0 & 1 & 0 & 0 & 0 & 0 & 0 & 0 \\
 0 & 0 & 0 & 0 & 0 & 0 & 1 & 0 \\
 0 & 0 & 0 & 0 & 0 & 0 & 0 & 1 \\
 0 & 0 & 0 & 0 & 1 & 0 & 0 & 0 \\
 0 & 0 & 0 & 0 & 0 & 1 & 0 & 0
\end{array}
\right),~~~~Y^{(8)}=\left(
\begin{array}{cccccccc}
 1 & 0 & 0 & 0 & 0 & 0 & 0 & 0 \\
 0 & 1 & 0 & 0 & 0 & 0 & 0 & 0 \\
 0 & 0 & 0 & 0 & 0 & 0 & 1 & 0 \\
 0 & 0 & 0 & 0 & 0 & 0 & 0 & 1 \\
 0 & 0 & 1 & 0 & 0 & 0 & 0 & 0 \\
 0 & 0 & 0 & 1 & 0 & 0 & 0 & 0 \\
 0 & 0 & 0 & 0 & 1 & 0 & 0 & 0 \\
 0 & 0 & 0 & 0 & 0 & 1 & 0 & 0
\end{array}
\right).\nonumber\end{eqnarray} The generators for $6$ are
\begin{eqnarray}A_1=A_2=A_3=A_4=I,\nonumber\end{eqnarray}where $I$ is the identity
matrix,
\begin{eqnarray}B_1^{(6)}=diag(1,-1,1,-1,1,-1),
~~~~B_2^{(6)}=diag(1,-1,-1,1,-1,1),\nonumber\end{eqnarray}
\begin{eqnarray}B_3^{(6)}=diag(-1,1,1,-1,-1,1),
~~~~B_4^{(6)}=diag(-1,1,-1,1,1,-1),\nonumber\end{eqnarray}
\begin{eqnarray}X^{(6)}=\left(
\begin{array}{cccccc}
 1 & 0 & 0 & 0 & 0 & 0 \\
 0 & 1 & 0 & 0 & 0 & 0 \\
 0 & 0 & 0 & 1 & 0 & 0 \\
 0 & 0 & 1 & 0 & 0 & 0 \\
 0 & 0 & 0 & 0 & 0 & 1 \\
 0 & 0 & 0 & 0 & 1 & 0
\end{array}
\right), ~~~~Y^{(6)}=\left(
\begin{array}{cccccc}
 0 & 0 & 0 & 0 & 1 & 0 \\
 0 & 0 & 0 & 0 & 0 & 1 \\
 1 & 0 & 0 & 0 & 0 & 0 \\
 0 & 1 & 0 & 0 & 0 & 0 \\
 0 & 0 & 1 & 0 & 0 & 0 \\
 0 & 0 & 0 & 1 & 0 & 0
\end{array}
\right).\nonumber\end{eqnarray} It can be checked directly that each
set of matrices respects Eqs. (\ref{group1})-(\ref{group4}).

\section{Appendix B}

In this appendix, we give the symmetry invariants which are used in
our model. These can be computed directly from the matrices given in
Appendix A.
\\ \\
\noindent{\bf (i) $\bm{16\times16}$ invariant} ($x_i,x'_j\sim16$):
\begin{eqnarray}f_1(x_i,x'_j)&=&x_1x'_{16}+x_2x'_{15}+x_3x'_{14}+x_4x'_{13}+x_5x'_{12}
+x_6x'_{11}+x_7x'_{10}+x_8x'_9\nonumber \\
&+& x_9x'_8+x_{10}x'_7+x_{11}x'_6+x_{12}x'_5+x_{13}x'_4+x_{14}x'_3
+x_{15}x'_2+x_{16}x'_1\nonumber\end{eqnarray}
\noindent {\bf (ii) $\bm{3\times3}$ invariant} ($t_i,t'_j\sim3$):
\begin{eqnarray}f_2(t_i,t'_j)=t_1t'_1+t_2t'_2+t_3t'_3\nonumber\end{eqnarray}\\
\noindent{\bf (iii) $\bm{48\times48}$ invariant} ($y_i,y'_j\sim48$):
\begin{eqnarray}f_3(y_i,y'_j)&=&y_1y'_{16}+y_2y'_{15}+y_3y'_{14}+y_4y'_{13}+y_5y'_{12}+y_6y'_{11}+y_7y'_{10}
+y_8y'_9\nonumber \\
&+& y_9y'_8+y_{10}y'_7+y_{11}y'_6+y_{12}y'_5+y_{13}y'_4+y_{14}y'_3
+y_{15}y'_2+y_{16}y'_1+y_{17}y'_{32}+y_{18}y'_{31}\nonumber\\
&+& y_{19}y'_{30}+y_{20}y'_{29}+y_{21}y'_{28}
+y_{22}y'_{27}+y_{23}y'_{26}+y_{24}y'_{25}+y_{25}y'_{24}+y_{26}y'_{23}+y_{27}y'_{22}
+y_{28}y'_{21}\nonumber \\
&+& y_{29}y'_{20}
+y_{30}y'_{19}+y_{31}y'_{18}+y_{32}y'_{17}+y_{33}y'_{48}+y_{34}y'_{47}+y_{35}y'_{46}+y_{36}y'_{45}+y_{37}y'_{44}
+y_{38}y'_{43}\nonumber\\
&+& y_{39}y'_{42}+y_{40}y'_{41}+y_{41}y'_{40}+y_{42}y'_{39}+y_{43}y'_{38}
+y_{44}y'_{37}+y_{45}y'_{36}+y_{46}y'_{35}+y_{47}y'_{34}+y_{48}y'_{33}\nonumber\end{eqnarray}
\noindent {\bf (iv) $\bm{8\times8}$ invariant} ($z_i,z'_j\sim8$):
\begin{eqnarray}f_4(z_i,z'_j)=z_1z'_2+z_2z'_1+z_3z'_4+z_4z'_3+z_5z'_6+z_6z'_5+z_7z'_8
+z_8z'_7\nonumber\end{eqnarray}
\noindent {\bf (v) $\bm{6\times6}$ invariant} ($w_i,w'_j\sim6$):
\begin{eqnarray}f_5(w_i,w'_j)=w_1w'_1+w_2w'_2+w_3w'_3+w_4w'_4+w_5w'_5+w_6w'_6\nonumber\end{eqnarray}
\noindent{\bf (vi) $\bm{16\times16\times16}$ invariant} ($x_i,x'_j,x''_k\sim16$):
\begin{eqnarray}g_1(x_i,x'_j,x''_k)&=&x_1x'_1x''_1+x_2x'_2x''_2+x_3x'_3x''_3+x_4x'_4x''_4+x_5x'_5x''_5+x_6x'_6x''_6
+x_7x'_7x''_7+x_8x'_8x''_8\nonumber \\
&+& x_9x'_9x''_9+x_{10}x'_{10}x''_{10}+x_{11}x'_{11}x''_{11}+x_{12}x'_{12}x''_{12}+x_{13}x'_{13}x''_{13}
+x_{14}x'_{14}x''_{14}+x_{15}x'_{15}x''_{15}\nonumber \\
&+& x_{16}x'_{16}x''_{16}\nonumber\end{eqnarray}
\noindent {\bf (vii) $\bm{3\times16\times48}$ invariant} ($t_i\sim3$, $x_j\sim16$,
$y_k\sim48$):
\begin{eqnarray}g_2(t_i,x_j,y_k)&=&t_1(x_{16}y_{33}+x_{15}y_{34}+x_8y_{35}+x_7y_{36}+x_{14}y_{37}+x_{13}y_{38}+x_6y_{39}+
x_5y_{40}\nonumber\\
&+&x_{12}y_{41}+x_{11}y_{42}+x_4y_{43}+x_3y_{44}+x_{10}y_{45}
+x_9y_{46}+x_2y_{47}+x_1y_{48})\nonumber \\
&+& t_2(x_{16}y_1+x_7y_{10}+x_6y_{11}+x_5y_{12}+x_4y_{13}+x_3y_{14}+x_2y_{15}+x_1y_{16}\nonumber\\
&+& x_{15}y_2+x_{14}y_3+x_{13}y_4+x_{12}y_5+x_{11}y_6+x_{10}y_7+x_9y_8+x_8y_9)\nonumber\\
&+& t_3(x_{16}y_{17}+x_{15}y_{18}+x_{12}y_{19}+x_{11}y_{20}+x_8y_{21}+x_7y_{22}+x_4y_{23}+
x_3y_{24}\nonumber\\
&+& x_{14}y_{25}+x_{13}y_{26}+x_{10}y_{27}+x_9y_{28}+x_6y_{29}+x_5y_{30}+x_2y_{31}+
x_1y_{32})\nonumber\end{eqnarray}
\noindent{\bf (viii) $\bm{8\times16\times48}$ invariant} ($z_i\sim8$, $x_j\sim16$,
$y_k\sim48$):
\begin{eqnarray}g_3(z_i,x_k,y_j)&=& z_1(x_{15}y_1+x_5y_{11}+x_3y_{13}+x_1y_{15}+x_{15}y_{17}+x_{11}y_{19}+x_7y_{21}+
x_3y_{23}
\nonumber \\
&+& x_{13}y_{25}+x_9y_{27}+x_5y_{29}+x_{13}y_3 +x_1y_{31}+x_{15}y_{33}+ x_7y_{35}+x_{13}y_{37}
\nonumber \\
&+& x_5y_{39}+x_{11}y_{41}+x_3y_{43}+x_9y_{45}+x_1y_{47}+x_{11}y_5+x_9y_7+x_7y_9)\nonumber\\
&+&z_2(x_8y_{10}+x_6y_{12}+x_4y_{14}+x_2y_{16}+x_{16}y_{18}+x_{16}y_2+x_{12}y_{20}
+ x_8y_{22}\nonumber \\
&+& x_4y_{24}+x_{14}y_{26}+x_{10}y_{28}+x_6y_{30} +x_2y_{32}+x_{16}y_{34}+x_8y_{36}
+x_{14}y_{38}\nonumber \\
&+& x_{14}y_4+x_6y_{40}+x_{12}y_{42}+x_4y_{44}+x_{10}y_{46}+x_2y_{48}+x_{12}y_6+x_{10}y_8)\nonumber\\
&+&z_3(x_{14}y_1+x_5y_{10}+x_2y_{13}+x_1y_{14}-x_{14}y_{17}-x_{13}y_{18}-x_{10}y_{19}
+ x_{13}y_2\nonumber \\
&-&x_9y_{20}-x_6y_{21}-x_5y_{22}-x_2y_{23}-x_1y_{24}-x_{14}y_{33}-x_{13}y_{34}-x_6y_{35}
\nonumber \\
&-&x_5y_{36}-x_{10}y_{41}-x_9y_{42}-x_2y_{43}-x_1y_{44}+x_{10}y_5+x_9y_6+x_6y_9)\nonumber\\
&+&z_4(x_8y_{11}+x_7y_{12}+x_4y_{15}+x_3y_{16}-x_{16}y_{25}-x_{15}y_{26}-x_{12}y_{27}-
x_{11}y_{28}\nonumber \\
&-& x_8y_{29}
+x_{16}y_3-x_7y_{30}-x_4y_{31}-x_3y_{32}-x_{16}y_{37}-x_{15}y_{38}-x_8y_{39}+x_{15}y_4\nonumber \\
&-&x_7y_{40}-x_{12}y_{45}-x_{11}y_{46}-x_4y_{47}-x_3y_{48}+x_{12}y_7+x_{11}y_8)\nonumber\\
&+&z_5(-x_{12}y_1-x_3y_{10}-x_2y_{11}-x_1y_{12}+x_{12}y_{17}+x_{11}y_{18}-x_{11}y_2+
x_4y_{21}+x_3y_{22}\nonumber \\
&+&x_{10}y_{25}+x_9y_{26}+x_2y_{29}-x_{10}y_3+x_1y_{30}-x_{12}y_{33}-x_{11}y_{34}-x_4y_{35}-x_3y_{36}
\nonumber \\
&-&x_{10}y_{37}-x_9y_{38}-x_2y_{39}-x_9y_4-x_1y_{40}-x_4y_9)
+z_6(-x_8y_{13}-x_7y_{14}-x_6y_{15}\nonumber \\
&-&x_5y_{16}+x_{16}y_{19}+x_{15}y_{20}+x_8y_{23}+
x_7y_{24}+x_{14}y_{27}+x_{13}y_{28}+x_6y_{31}+x_5y_{32}\nonumber\\
&-&x_{16}y_{41}-x_{15}y_{42}-x_8y_{43}-x_7y_{44}-x_{14}y_{45}-x_{13}y_{46}-x_6y_{47}
-x_5y_{48}-x_{16}y_5 \nonumber \\
&-&x_{15}y_6-x_{14}y_7-x_{13}y_8)
+ z_7(-x_8y_1-x_8y_{17}-x_7y_{18}-x_4y_{19}-x_7y_2-x_3y_{20}\nonumber \\
&-&x_6y_{25}-x_5y_{26}
-x_2y_{27}-x_1y_{28}-x_6y_3+x_8y_{33}
+x_7y_{34}+x_6y_{37}+x_5y_{38}-x_5y_4
\nonumber \\
&+&x_4y_{41} + x_3y_{42}+x_2y_{45}
+ x_1y_{46}-x_4y_5-x_3y_6-x_2y_7-x_1y_8) \nonumber \\
&+& z_8(x_{15}y_{36}-x_{14}y_{11}-x_{13}y_{12}-x_{12}y_{13}-x_{11}y_{14}-x_{10}y_{15}-x_9y_{16}-
x_{16}y_{21}\nonumber \\
&-&x_{15}y_{22}-x_{12}y_{23}-x_{11}y_{24}-x_{14}y_{29}
-x_{13}y_{30}-x_{10}y_{31}-x_9y_{32}\nonumber \\
&+&x_{16}y_{35}-x_{15}y_{10}+x_{14}y_{39}+x_{13}y_{40}+
x_{12}y_{43}+x_{11}y_{44}+x_{10}y_{47}+x_9y_{48}-x_{16}y_9)\nonumber
\end{eqnarray}
\noindent{\bf (ix) $\bm{16\times48\times48}$ invariant} ($x_i\sim16$; $y_j,y'_k\sim48$):
\begin{eqnarray}g_4(x_i,y_j,y'_k)&=&x_1y_1y'_1+x_2y_2y'_2+x_3y_3y'_3+x_4y_4y'_4+x_5y_5y'_5+x_6y_6y'_6+x_7y_7y'_7+
x_8y_8y'_8\nonumber \\
&+& x_9y_9y'_9+x_{10}y_{10}y'_{10}+x_{11}y_{11}y'_{11}+x_{12}y_{12}y'_{12}+x_{13}y_{13}y'_{13}
+x_{14}y_{14}y'_{14}+x_{15}y_{15}y'_{15}\nonumber \\
&+& x_{16}y_{16}y'_{16}
+ x_1y_{17}y'_{17}+
x_2y_{18}y'_{18}+x_3y_{25}y'_{25}+x_4y_{26}y'_{26}+x_5y_{19}y'_{19}+x_6y_{20}y'_{20}+x_7y_{27}y'_{27}
\nonumber \\
&+&x_8y_{28}y'_{28}
+ x_9y_{21}y'_{21}+x_{10}y_{22}y'_{22}+x_{11}y_{29}y'_{29}+x_{12}y_{30}y'_{30}
+x_{13}y_{23}y'_{23}+x_{14}y_{24}y'_{24}
\nonumber \\
&+& x_{15}y_{31}y'_{31} + x_{16}y_{32}y'_{32}+ x_1y_{33}y'_{33}
+x_2y_{34}y'_{34}+x_3y_{37}y'_{37}+x_4y_{38}y'_{38}
+ x_5y_{41}y'_{41}\nonumber \\
&+& x_6y_{42}y'_{42}+x_7y_{45}y'_{45}
+x_8y_{46}y'_{46}
+ x_9y_{35}y'_{35}+x_{10}y_{36}y'_{36}+x_{11}y_{39}y'_{39}+x_{12}y_{40}y'_{40}
\nonumber \\
&+&x_{13}y_{43}y'_{43}
+ x_{14}y_{44}y'_{44}+x_{15}y_{47}y'_{47}+x_{16}y_{48}y'_{48}\nonumber\end{eqnarray}
\noindent{\bf (x) $\bm{1'\times3\times3}$ invariant} ($s'\sim1'$; $t_i,t'_j\sim3$):
\begin{eqnarray}g_5(s',t_i,t'_j)=s'(t_1t'_1+\omega^2t_2t'_2+\omega t_3t'_3)\nonumber
\end{eqnarray}
\noindent{\bf (xi) $\bm{1''\times3\times3}$ invariant} ($s''\sim1''$; $t_i,t'_j\sim3$):
\begin{eqnarray}g_6(s'',t_i,t_j)=s''(t_1t'_1+\omega t_2t'_2+\omega^2t_3t'_3)\nonumber\end{eqnarray}\\
\noindent{\bf (xii) $\bm{6\times16\times16}$ invariant} ($w_i\sim6$; $x_j,x'_k\sim16$):
\begin{eqnarray}g_7(w_i,x_j,x'_k)&=&w_1(x_1x'_{16}+x_2x'_{15}+x_3x'_{14}+x_4x'_{13}-x_5x'_{12}
-x_6x'_{11}-x_7x'_{10}-x_8x'_9\nonumber\\
&-&x_9x'_8-x_{10}x'_7-x_{11}x'_6-x_{12}x'_5+x_{13}x'_4+x_{14}x'_3
+x_{15}x'_2+x_{16}x'_1)\nonumber\\
&+& w_2(x_1x'_{16}-x_2x'_{15}-x_3x'_{14}+x_4x'_{13}+x_5x'_{12}
-x_6x'_{11}-x_7x'_{10}+x_8x'_9\nonumber\\
&+&x_9x'_8-x_{10}x'_7-x_{11}x'_6+x_{12}x'_5+x_{13}x'_4-x_{14}x'_3
-x_{15}x'_2+x_{16}x'_1)\nonumber\\
&+& w_3(x_1x'_{16}+x_2x'_{15}-x_3x'_{14}-x_4x'_{13}+x_5x'_{12}
+x_6x'_{11}-x_7x'_{10}-x_8x'_9\nonumber\\
&-&x_9x'_8-x_{10}x'_7+x_{11}x'_6+x_{12}x'_5-x_{13}x'_4-x_{14}x'_3
+x_{15}x'_2+x_{16}x'_1)\nonumber\\
&+&w_4(x_1x'_{16}-x_2x'_{15}+x_3x'_{14}-x_4x'_{13}-x_5x'_{12}
+x_6x'_{11}-x_7x'_{10}+x_8x'_9\nonumber\\
&+&x_9x'_8-x_{10}x'_7+x_{11}x'_6-x_{12}x'_5-x_{13}x'_4+x_{14}x'_3
-x_{15}x'_2+x_{16}x'_1)\nonumber\\
&+& w_5(x_1x'_{16}+x_2x'_{15}-x_3x'_{14}-x_4x'_{13}-x_5x'_{12}
-x_6x'_{11}+x_7x'_{10}+x_8x'_9\nonumber\\
&+&x_9x'_8-x_{10}x'_7-x_{11}x'_6-x_{12}x'_5-x_{13}x'_4-x_{14}x'_3
+x_{15}x'_2+x_{16}x'_1)\nonumber\\
&+& w_6(x_1x'_{16}-x_2x'_{15}+x_3x'_{14}-x_4x'_{13}+x_5x'_{12}
-x_6x'_{11}+x_7x'_{10}-x_8x'_9\nonumber\\
&-&x_9x'_8+x_{10}x'_7-x_{11}x'_6+x_{12}x'_5-x_{13}x'_4+x_{14}x'_3
-x_{15}x'_2+x_{16}x'_1)\nonumber\end{eqnarray}
\noindent{\bf (xiii)
$\bm{6\times16\times16'}$ invariant} ($w_i\sim6$; $x_j,x'_k\sim16'$):
\begin{eqnarray}g_8(w_i,x_j,x'_k)&=&w_1(x_1x'_{16}+x_2x'_{15}+x_3x'_{14}+x_4x'_{13}-x_5x'_{12}
-x_6x'_{11}-x_7x'_{10}-x_8x'_9\nonumber\\
&-&x_9x'_8-x_{10}x'_7-x_{11}x'_6-x_{12}x'_5+x_{13}x'_4+x_{14}x'_3
+x_{15}x'_2+x_{16}x'_1)\nonumber\\
&+&w_2(x_1x'_{16}-x_2x'_{15}-x_3x'_{14}+x_4x'_{13}+x_5x'_{12}
-x_6x'_{11}-x_7x'_{10}+x_8x'_9\nonumber\\
&+&x_9x'_8-x_{10}x'_7-x_{11}x'_6+x_{12}x'_5+x_{13}x'_4-x_{14}x'_3
-x_{15}x'_2+x_{16}x'_1)\nonumber\\
&+& \omega^2w_3(x_1x'_{16}+x_2x'_{15}-x_3x'_{14}-x_4x'_{13}+x_5x'_{12}
+x_6x'_{11}-x_7x'_{10}-x_8x'_9\nonumber\\
&-&x_9x'_8-x_{10}x'_7+x_{11}x'_6+x_{12}x'_5-x_{13}x'_4-x_{14}x'_3
+x_{15}x'_2+x_{16}x'_1)\nonumber\\
&+& \omega^2w_4(x_1x'_{16}-x_2x'_{15}+x_3x'_{14}-x_4x'_{13}-x_5x'_{12}
+x_6x'_{11}-x_7x'_{10}+x_8x'_9\nonumber\\
&+&x_9x'_8-x_{10}x'_7+x_{11}x'_6-x_{12}x'_5-x_{13}x'_4+x_{14}x'_3
-x_{15}x'_2+x_{16}x'_1)\nonumber\\
&+& \omega w_5(x_1x'_{16}+x_2x'_{15}-x_3x'_{14}-x_4x'_{13}-x_5x'_{12}
-x_6x'_{11}+x_7x'_{10}+x_8x'_9\nonumber\\
&+&x_9x'_8-x_{10}x'_7-x_{11}x'_6-x_{12}x'_5-x_{13}x'_4-x_{14}x'_3
+x_{15}x'_2+x_{16}x'_1)\nonumber\\
&+& \omega w_6(x_1x'_{16}-x_2x'_{15}+x_3x'_{14}-x_4x'_{13}+x_5x'_{12}
-x_6x'_{11}+x_7x'_{10}-x_8x'_9\nonumber\\
&-&x_9x'_8+x_{10}x'_7-x_{11}x'_6+x_{12}x'_5-x_{13}x'_4+x_{14}x'_3
-x_{15}x'_2+x_{16}x'_1)\nonumber\end{eqnarray}
\noindent{\bf (xiv) $\bm{6\times16\times16''}$
invariant} ($w_i\sim6$; $x_j,x'_k\sim16''$):
\begin{eqnarray}g_9(w_i,x_j,x'_k)&=&w_1(x_1x'_{16}+x_2x'_{15}+x_3x'_{14}+x_4x'_{13}-x_5x'_{12}
-x_6x'_{11}-x_7x'_{10}-x_8x'_9\nonumber\\
&-&x_9x'_8-x_{10}x'_7-x_{11}x'_6-x_{12}x'_5+x_{13}x'_4+x_{14}x'_3
+x_{15}x'_2+x_{16}x'_1)\nonumber\\
&+& w_2(x_1x'_{16}-x_2x'_{15}-x_3x'_{14}+x_4x'_{13}+x_5x'_{12}
-x_6x'_{11}-x_7x'_{10}+x_8x'_9\nonumber\\
&+&x_9x'_8-x_{10}x'_7-x_{11}x'_6+x_{12}x'_5+x_{13}x'_4-x_{14}x'_3
-x_{15}x'_2+x_{16}x'_1)\nonumber\\
&+& \omega w_3(x_1x'_{16}+x_2x'_{15}-x_3x'_{14}-x_4x'_{13}+x_5x'_{12}
+x_6x'_{11}-x_7x'_{10}-x_8x'_9\nonumber\\
&-&x_9x'_8-x_{10}x'_7+x_{11}x'_6+x_{12}x'_5-x_{13}x'_4-x_{14}x'_3
+x_{15}x'_2+x_{16}x'_1)\nonumber\\
&+& \omega w_4(x_1x'_{16}-x_2x'_{15}+x_3x'_{14}-x_4x'_{13}-x_5x'_{12}
+x_6x'_{11}-x_7x'_{10}+x_8x'_9\nonumber\\
&+&x_9x'_8-x_{10}x'_7+x_{11}x'_6-x_{12}x'_5-x_{13}x'_4+x_{14}x'_3
-x_{15}x'_2+x_{16}x'_1)\nonumber\\
&+&\omega^2w_5(x_1x'_{16}+x_2x'_{15}-x_3x'_{14}-x_4x'_{13}-x_5x'_{12}
-x_6x'_{11}+x_7x'_{10}+x_8x'_9\nonumber\\
&+&x_9x'_8-x_{10}x'_7-x_{11}x'_6-x_{12}x'_5-x_{13}x'_4-x_{14}x'_3
+x_{15}x'_2+x_{16}x'_1)\nonumber\\
&+& \omega^2w_6(x_1x'_{16}-x_2x'_{15}+x_3x'_{14}-x_4x'_{13}+x_5x'_{12}
-x_6x'_{11}+x_7x'_{10}-x_8x'_9\nonumber\\
&-&x_9x'_8+x_{10}x'_7-x_{11}x'_6+x_{12}x'_5-x_{13}x'_4+x_{14}x'_3
-x_{15}x'_2+x_{16}x'_1)\nonumber\end{eqnarray}

For our purposes, it
suffices to have the $16\times16\times16\times16$ and
$3\times3\times3\times3$ invariants for the case where all
four fields are the same.
\vspace*{0.1in}

\noindent{\bf (xv) $\bm{16\times16\times16\times16}$ invariants}
($x_i\sim16$):
\begin{eqnarray}h_1(x_i)&=&x_1^2x_{16}^2+x_2^2x_{15}^2+x_3^2x_{14}^2+x_4^2x_{13}^2
+x_5^2x_{12}^2+x_6^2x_{11}^2+x_7^2x_{10}^2+x_8^2x_9^2,\nonumber\\
h_2(x_i)&=&x_1x_2x_{15}x_{16}+x_1x_3x_{14}x_{16}+x_2x_4x_{13}x_{15}
+x_3x_4x_{13}x_{14}+x_1x_5x_{12}x_{16}+x_4x_5x_{12}x_{13}
\nonumber\\
&+&x_2x_6x_{11}x_{15}+x_3x_6x_{11}x_{14}+x_5x_6x_{11}x_{12}
+x_2x_7x_{10}x_{15}+x_3x_7x_{10}x_{14}+x_5x_7x_{10}x_{12}
\nonumber\\
&+&x_1x_8x_9x_{16}+x_4x_8x_9x_{13}+x_6x_8x_9x_{11}+x_7x_8x_9x_{10}\nonumber\end{eqnarray}
\noindent{\bf (xvi)
$\bm{3\times3\times3\times3}$ invariants} ($w_i\sim3$):
\begin{eqnarray}h_3(w_i)&=&w_1^4+w_2^4+w_3^4,
\nonumber\\
h_4(w_i)&=&w_1^2w_2^2+w_1^2w_3^2+w_2^2w_3^2,\nonumber\end{eqnarray}

\section{Appendix C}

In this appendix, we show how the light charged lepton mass matrix is
computed. In the basis with
$(\overline{e}_{L1},\overline{e}_{L2},\overline{e}_{L3},\overline{E}_{L1},\overline{E}_{L2},\overline{E}_{L3})$
on the left and $(e_{R1},e_{R2},e_{R3},E_{R1},E_{R2},E_{R3})$ on the
right, the mass matrix has the form
\begin{eqnarray}{\cal M}_e=\left(
                  \begin{array}{cc}
                    0 & M' \\
                    m & M \\
                  \end{array}
                \right),\nonumber
\end{eqnarray}
with
\begin{eqnarray}m=\left(
                    \begin{array}{ccc}
                      \kappa v & 0 & 0 \\
                      0 & \kappa v & 0 \\
                      0 & 0 & \kappa v \\
                    \end{array}
                  \right),
\nonumber\end{eqnarray}
\begin{eqnarray}M=\left(
                    \begin{array}{ccc}
                      m_E & 0 & 0 \\
                      0 & m_E & 0 \\
                      0 & 0 & m_E \\
                    \end{array}
                  \right),
\nonumber\end{eqnarray} and
\begin{eqnarray}M'=\left(
                     \begin{array}{ccc}
                       \epsilon_1v_{\chi} & \epsilon_1v_{\chi} & \epsilon_1v_{\chi} \\
                       \epsilon_2v_{\chi} & \omega^2\epsilon_2v_{\chi} & \omega\epsilon_2v_{\chi} \\
                       \epsilon_3v_{\chi} & \omega\epsilon_3v_{\chi} & \omega^2\epsilon_3v_{\chi} \\
                     \end{array}
                   \right)=\left(
                             \begin{array}{ccc}
                               \epsilon_1 & 0 & 0 \\
                               0 & \epsilon_2 & 0 \\
                               0 & 0 & \epsilon_3 \\
                             \end{array}
                           \right)\left(
                                    \begin{array}{ccc}
                                      1 & 1 & 1 \\
                                      1 & \omega^2 & \omega \\
                                      1 & \omega & \omega^2 \\
                                    \end{array}
                                  \right).
\nonumber\end{eqnarray} Here, $m$ only contains entries at the EW
scale, while $M$ and $M'$ contain entries at the higher scale $M_*$.
To order $M_W^2/M_*^2$, the left-handed mass-squared matrix ${\cal
M}_e^{\dag}{\cal M}_e$ is block-diagonalized by \cite{bb}
\begin{eqnarray}{\cal U}_L=\left(
                  \begin{array}{cc}
                    I & m^{\dag}M(M^{\dag}M+M'^{\dag}M')^{-1} \\
                    (M^{\dag}M+M'^{\dag}M')^{-1}M^{\dag}m & I \\
                  \end{array}
                \right).\nonumber
\end{eqnarray} The upper left entry of ${\cal U}_L{\cal M}_e^{\dag}{\cal M}_e{\cal
U}_L^{\dag}$ is the light left-handed mass-squared matrix
\begin{eqnarray}M_e^{\dag}M_e=m^{\dag}m-m^{\dag}M(M^{\dag}M+M'^{\dag}M')^{-1}M^{\dag}m\nonumber\end{eqnarray}
{\small\begin{eqnarray}=\frac{|\kappa vm_E|^2}{3}\left(
                                     \begin{array}{ccc}
                                       1 & 1 & 1 \\
                                       1 & \omega & \omega^2 \\
                                       1 & \omega^2 & \omega \\
                                     \end{array}
                                   \right)\left(
                                            \begin{array}{ccc}
                                              |m_E|^2+3|\epsilon_1v_{\chi}|^2 & 0 & 0 \\
                                              0 & |m_E|^2+3|\epsilon_2v_{\chi}|^2 & 0 \\
                                              0 & 0 & |m_E|^2+3|\epsilon_3v_{\chi}|^2 \\
                                            \end{array}
                                          \right)^{-1}\left(
                                     \begin{array}{ccc}
                                       1 & 1 & 1 \\
                                       1 & \omega^2 & \omega \\
                                       1 & \omega & \omega^2 \\
                                     \end{array}
                                   \right).\nonumber\end{eqnarray}}
This yields the masses given in (\ref{me}).

\section{Appendix D}

In this appendix, we show how the neutrino mass matrix is computed.
From Appendix B, the term in Eq. (\ref{Lnu}) that mixes $N$ and $N'$ is
\begin{eqnarray}g_3(N,\langle\phi\rangle,N')&=&v_{\phi}N_1(N'_{35}+N'_{36}+N'_{39}+N'_{40}+N'_{41}+N'_{42}
+N'_{45}+N'_{46})\nonumber\\
&+& v_{\phi}N_2(N'_{5}+N'_{6}+N'_{7}+N'_{8}+N'_{9}+N'_{10}+N'_{11}+N'_{12})\nonumber\\
&+&v_{\phi}N_3(N'_{19}+N'_{20}+N'_{21}+N'_{22}+N'_{27}+N'_{28}+N'_{29}+N'_{30}).\nonumber\end{eqnarray}
The term that mixes $N'$ and $N''$ is
\begin{eqnarray}g_4(N'',\langle\phi\rangle,N')&=&v_{\phi}N''_1(N'_{5}+N'_{7}+N'_{9}+N'_{11})+v_{\phi}N''_2(N'_{6}+N'_{8}
+N'_{10}+N'_{12})\nonumber\\
&+&v_{\phi}N''_3(N'_{5}+N'_{6}+N'_{9}+N'_{10})+v_{\phi}N''_4(N'_{7}+N'_{8}+N'_{11}+N'_{12})\nonumber\\
&-&v_{\phi}N''_5(N'_{1}+N'_{2}+N'_{3}+N'_{4})-v_{\phi}N''_6(N'_{13}+N'_{14}+N'_{15}+N'_{16})\nonumber\\
&-&v_{\phi}N''_7(N'_{1}+N'_{2}+N'_{3}+N'_{4})-v_{\phi}N''_8(N'_{13}+N'_{14}+N'_{15}+N'_{16})\nonumber\\
&+&v_{\phi}N''_1(N'_{19}+N'_{21}+N'_{27}+N'_{29})+v_{\phi}N''_2(N'_{20}+N'_{22}+N'_{28}+N'_{30})\nonumber\\
&-&v_{\phi}N''_3(N'_{19}+N'_{20}+N'_{21}+N'_{22})-v_{\phi}N''_4(N'_{27}+N'_{28}+N'_{29}+N'_{30})\nonumber\\
&+&v_{\phi}N''_5(N'_{17}+N'_{18}+N'_{25}+N'_{26})+v_{\phi}N''_6(N'_{23}+N'_{29}+N'_{31}+N'_{32})\nonumber\\
&-&v_{\phi}N''_7(N'_{17}+N'_{18}+N'_{25}+N'_{26})-v_{\phi}N''_8(N'_{23}+N'_{29}+N'_{31}+N'_{32})\nonumber\\
&+&v_{\phi}N''_1(N'_{35}+N'_{39}+N'_{41}+N'_{45})+v_{\phi}N''_2(N'_{36}+N'_{40}+N'_{42}+N'_{46})\nonumber\\
&-&v_{\phi}N''_3(N'_{35}+N'_{36}+N'_{41}+N'_{42})-v_{\phi}N''_4(N'_{39}+N'_{40}+N'_{45}+N'_{46})\nonumber\\
&+&v_{\phi}N''_5(N'_{33}+N'_{34}+N'_{37}+N'_{38})+v_{\phi}N''_6(N'_{43}+N'_{44}+N'_{47}+N'_{48})\nonumber\\
&-&v_{\phi}N''_7(N'_{33}+N'_{34}+N'_{37}+N'_{38})-v_{\phi}N''_8(N'_{43}+N'_{44}+N'_{47}+N'_{48}).\nonumber\end{eqnarray}
Since the symmetries $B_1$, $B_2$, $B_3B_4$, and $A_3A_4$ are
unbroken, components of $N'$ and $N''$ that transform under these
symmetries cannot mix with the light neutrinos. This leaves
\begin{eqnarray}p_1&=&\frac{N'_{5}+N'_{6}+N'_{7}+N'_{8}+N'_{9}+N'_{10}+N'_{11}+N'_{12}}{\sqrt{8}},\nonumber
\\
p_2&=&\frac{N'_{19}+N'_{20}+N'_{21}+N'_{22}+N'_{27}+N'_{28}+N'_{29}+N'_{30}}{\sqrt{8}},\nonumber\\
p_3&=&\frac{N'_{35}+N'_{36}+N'_{39}+N'_{40}+N'_{41}+N'_{42}+N'_{45}+N'_{46}}{\sqrt{8}},\nonumber\\
q_1&=&\frac{N''_{1}+N''_{2}}{\sqrt{2}},~~~~q_2=\frac{N''_{3}+N''_{4}}{\sqrt{2}}.\nonumber\end{eqnarray}
We now have
\begin{eqnarray}g_3(N,\langle\phi\rangle,N')&=&\sqrt{8}v_{\phi}(N_1p_3+N_2p_1+N_3p_2),\nonumber\\
g_4(N'',\langle\phi\rangle,N')&=&2v_{\phi}(q_1p_1+q_2p_1+q_1p_2-q_2p_2+q_1p_3-q_2p_3)+...,\nonumber\end{eqnarray}
where the ellipses in the second equation refer to terms involving only decoupled components. The mass matrix for
$(\nu_1,\nu_2,\nu_3,N_1,N_2,N_3,p_1,p_2,p_3,q_1,q_2)$ has the form
\begin{eqnarray}{\cal M}_{\nu}=\left(
                  \begin{array}{cc}
                    0 & m \\
                    m^T & M \\
                  \end{array}
                \right),
\nonumber\end{eqnarray}
with
\begin{eqnarray}m=\left(
                  \begin{array}{cccccccc}
                    \frac{1}{2}\lambda v & 0 & 0 & 0 & 0 & 0 & 0 & 0 \\
                    0 & \frac{1}{2}\lambda v & 0 & 0 & 0 & 0 & 0 & 0 \\
                    0 & 0 & \frac{1}{2}\lambda v & 0 & 0 & 0 & 0 & 0 \\
                  \end{array}
                \right)
,
\nonumber\end{eqnarray}
\begin{eqnarray}M=\left(
                    \begin{array}{cccccccc}
                      m_N & 0 & 0 & 0 & 0 & \sqrt{2}\alpha_1v_{\phi} & 0 & 0 \\
                      0 & m_N & 0 & \sqrt{2}\alpha_1v_{\phi} & 0 & 0 & 0 & 0 \\
                      0 & 0 & m_N & 0 & \sqrt{2}\alpha_1v_{\phi} & 0 & 0 & 0 \\
                      0 & \sqrt{2}\alpha_1v_{\phi} & 0 & m'_N+\beta v_{\phi} & 0 & 0 & \alpha_2v_{\phi} & \alpha_2v_{\phi} \\
                      0 & 0 & \sqrt{2}\alpha_1v_{\phi} & 0 & m'_N+\beta v_{\phi} & 0 & \alpha_2v_{\phi} & -\alpha_2v_{\phi} \\
                      \sqrt{2}\alpha_1v_{\phi} & 0 & 0 & 0 & 0 & m'_N+\beta v_{\phi} & \alpha_2v_{\phi} & -\alpha_2v_{\phi} \\
                      0 & 0 & 0 & \alpha_2v_{\phi} & \alpha_2v_{\phi} & \alpha_2v_{\phi} & m''_N & 0 \\
                      0 & 0 & 0 & \alpha_2v_{\phi} & -\alpha_2v_{\phi} & -\alpha_2v_{\phi} & 0 & m''_N \\
                    \end{array}
                  \right)
.
\nonumber\end{eqnarray}
Here, $m$ only contains entries at the EW scale, while $M$ contains entries at the higher scale $M_*$.
To order $M_W^2/M_*^2$, ${\cal M}_{\nu}$ is block-diagonalized by
\begin{eqnarray}{\cal U}_{\nu}=\left(
                  \begin{array}{cc}
                    I & -mM^{-1} \\
                    M^{-1}m^T & I \\
                  \end{array}
                \right).
\nonumber\end{eqnarray}
The light neutrino mass matrix $M_{\nu}$ is given by the upper-left block of ${\cal U}_{\nu}{\cal M}_{\nu}{\cal U}_{\nu}^T$,
\begin{eqnarray}M_{\nu}=-mM^{-1}m^T.
\nonumber\end{eqnarray}
Let
\begin{eqnarray}S=\frac{1}{\sqrt{2}}\left(
                    \begin{array}{cccccccc}
                      1 & 0 & 0 & 0 & 0 & 0 & 1 & 0 \\
                      0 & 0 & 0 & \sqrt{2} & 0 & 0 & 0 & 0 \\
                      1 & 0 & 0 & 0 & 0 & 0 & -1 & 0 \\
                      0 & 0 & 0 & 0 & \sqrt{2} & 0 & 0 & 0 \\
                      0 & 1 & 0 & 0 & 0 & 0 & 0 & 1 \\
                      0 & 1 & 0 & 0 & 0 & 0 & 0 & -1 \\
                      0 & 0 & 1 & 0 & 0 & 1 & 0 & 0 \\
                      0 & 0 & -1 & 0 & 0 & 1 & 0 & 0 \\
                    \end{array}
                  \right).
\nonumber\end{eqnarray}
Then,
\begin{eqnarray}S^{-1}MS=\left(
                           \begin{array}{ccc}
                             A & 0 & 0 \\
                             0 & B & 0 \\
                             0 & 0 & C \\
                           \end{array}
                         \right),
\nonumber\end{eqnarray}
with
\begin{eqnarray}A=\left(
                           \begin{array}{ccc}
                             m_N & \sqrt{2}\alpha_1v_{\phi} & 0 \\
                             \sqrt{2}\alpha_1v_{\phi} & m'_N+\beta v_{\phi} & 2\alpha_2v_{\phi} \\
                             0 & 2\alpha_2v_{\phi} & m''_N \\
                           \end{array}
                         \right),
\nonumber\end{eqnarray}
\begin{eqnarray}B=\left(
                           \begin{array}{ccc}
                             m_N & \sqrt{2}\alpha_1v_{\phi} & 0 \\
                             \sqrt{2}\alpha_1v_{\phi} & m'_N+\beta v_{\phi} & \sqrt{2}\alpha_2v_{\phi} \\
                             0 & \sqrt{2}\alpha_2v_{\phi} & m''_N \\
                           \end{array}
                         \right),
\nonumber\end{eqnarray}
\begin{eqnarray}C=\left(
                          \begin{array}{cc}
                            m_N & -\sqrt{2}\alpha_1v_{\phi} \\
                            -\sqrt{2}\alpha_1v_{\phi} & m'_N+\beta v_{\phi} \\
                          \end{array}
                        \right).
\nonumber\end{eqnarray}
So, we can write
\begin{eqnarray}M_{\nu}=-mS\left(
                           \begin{array}{ccc}
                             A^{-1} & 0 & 0 \\
                             0 & B^{-1} & 0 \\
                             0 & 0 & C^{-1} \\
                           \end{array}
                         \right)S^{-1}m^T
\nonumber\end{eqnarray}
\begin{eqnarray}=-\frac{\lambda^2v^2}{8}\left(
                                            \begin{array}{ccc}
                                              (A^{-1})_{11}+(C^{-1})_{11} & 0 & (A^{-1})_{11}-(C^{-1})_{11} \\
                                              0 & 2(B^{-1})_{11} & 0 \\
                                              (A^{-1})_{11}-(C^{-1})_{11} & 0 & (A^{-1})_{11}+(C^{-1})_{11} \\
                                            \end{array}
                                          \right).
\nonumber\end{eqnarray} This mass matrix is diagonalized by (\ref{Unu}),
and the masses are given by
\begin{eqnarray}m_1=\left|\frac{\lambda^2v^2}{4}(A^{-1})_{11}\right|,~~m_2=\left|\frac{\lambda^2v^2}{4}(B^{-1})_{11}\right|,~~
m_3=\left|\frac{\lambda^2v^2}{4}(C^{-1})_{11}\right|.
\nonumber\end{eqnarray}

\section*{Acknowledgments}  This work is supported in part by Department of Energy Grant Numbers
DE-FG02-04ER41306 and DE-FG02-ER46140.

\end{document}